\documentclass[pdflatex,sn-basic,iicol]{sn-jnl}

\usepackage{multirow}%
\usepackage{amsmath,amssymb,amsfonts}%
\usepackage{amsthm}%
\usepackage{mathrsfs}%
\usepackage[title]{appendix}%
\usepackage{xcolor}%
\usepackage{textcomp}%
\usepackage{manyfoot}%
\usepackage{booktabs}%
\usepackage{algorithm}%
\usepackage[utf8]{inputenc}%
\usepackage{algorithmicx}%
\usepackage{algpseudocode}%
\usepackage{listings}%
\usepackage{graphicx}%
\usepackage{pifont}
\usepackage{array}
\usepackage{caption}
\usepackage{pifont}
\usepackage{fontawesome} 

\newcommand{\rev}[1]{
    \ifthenelse{\equal{#1}{3}}{
        \faStar \faStar \faStar
    }{
        \ifthenelse{\equal{#1}{2.5}}{
            \faStar \faStar \faStarHalfO
        }{
            \ifthenelse{\equal{#1}{2}}{
                \faStar \faStar \faStarO
            }{
                \ifthenelse{\equal{#1}{1.5}}{
                    \faStar \faStarHalfO \faStarO
                }{
                    \ifthenelse{\equal{#1}{1}}{
                        \faStar \faStarO \faStarO
                    }{
                        \faStarO \faStarO \faStarO
                    }
                }
            }
        }
    }
}

\theoremstyle{thmstyleone}%
\theoremstyle{thmstyletwo}%

\theoremstyle{thmstylethree}%

\begin{document}

\title[Article Title]{Controllable Video Generation: A Survey}



\author[1]{\fnm{Yue} \sur{Ma}}\email{mayuefighting@gmail.com}
\equalcont{These authors contributed equally to this work.}
\author[2]{\fnm{Kunyu} \sur{Feng}}\email{fengkunyu513@gmail.com}
\equalcont{These authors contributed equally to this work.}

\author[3]{\fnm{Zhongyuan} \sur{Hu}}\email{huzhongyyuan@gmail.com}
\equalcont{These authors contributed equally to this work.}

\author[3]{\fnm{Xinyu} \sur{Wang}}\email{cpwxyxwcp@gmail.com}
\equalcont{These authors contributed equally to this work.}
\author[1]{\fnm{Yucheng} \sur{Wang}}\email{ywangls@connect.ust.hk}

\author[1]{\fnm{Mingzhe} \sur{Zheng}}\email{mzhengar@connect.ust.hk}

\author[2]{\fnm{Bingyuan} \sur{Wang}}\email{bwang667@connect.hkust-gz.edu.cn}

\author[4]{\fnm{Qinghe} \sur{Wang}}\email{qinghewang@mail.dlut.edu.cn}

\author[1]{\fnm{Xuanhua} \sur{He}}\email{xhecd@connect.ust.hk}

\author[3]{\fnm{Hongfa} \sur{Wang}}\email{wanghf22@mails.tsinghua.edu.cn}

\author[3]{\fnm{Chenyang} \sur{Zhu}}\email{chenyangzhu.cs@gmail.com}

\author[1]{\fnm{Hongyu} \sur{Liu}}\email{hliudq@connect.ust.hk}

\author[1]{\fnm{Yingqing} \sur{He}}\email{yhebm@connect.ust.hk}

\author[1,2]{\fnm{Zeyu} \sur{Wang}}\email{zeyuwang@ust.hk}

\author[6]{\fnm{Zhifeng} \sur{Li}}\email{zhifeng0.li@gmail.com}

\author[3]{\fnm{Xiu} \sur{Li}}\email{li.xiu@sz.tsinghua.edu.cn}

\author[1]{\fnm{Sirui} \sur{Han}}\email{siruihan@ust.hk}

\author[1]{\fnm{Yike} \sur{Guo}}\email{yikeguo@ust.hk}

\author[6]{\fnm{Wei} \sur{Liu}}\email{wl2223@columbia.edu}

\author[1]{\fnm{Dan} \sur{Xu}}\email{danxu@cse.ust.hk}

\author*[5]{\fnm{Linfeng} \sur{Zhang}}\email{zhanglinfeng@sjtu.edu.cn}

\author*[1]{\fnm{Qifeng} \sur{Chen}}\email{cqf@ust.hk}

\affil[1]{\orgname{Hong Kong University of Science and Technology}, \orgaddress{\city{Hong Kong}, \country{China}}}

\affil[2]{\orgname{Hong Kong University of Science and Technology(Guang Zhou)}, \orgaddress{\city{Guang Zhou}, \country{China}}}

\affil[3]{\orgname{Tsinghua University}, \orgaddress{\city{Shenzhen}, \country{China}}}

\affil[4]{ \orgname{Dalian University of Technology}, \orgaddress{\city{Dalian}, \country{China}}}

\affil[5]{\orgdiv{EPIC Lab of Artificial Intelligence}, \orgname{Shanghai Jiao Tong University}, \orgaddress{\city{Shanghai}, \country{China}}}

\affil[6]{\orgname{Tencent}, \orgaddress{\city{Shenzhen}, \country{China}}}


\abstract{With the rapid development of AI-generated content (AIGC), video generation has emerged as one of its most dynamic and impactful subfields. In particular, the advancement of video generation foundation models has led to growing demand for controllable video generation methods that can more accurately reflect user intent. Most existing foundation models are designed for text-to-video generation, where text prompts alone are often insufficient to express complex, multi-modal, and fine-grained user requirements. This limitation makes it challenging for users to generate videos with precise control using current models. To address this issue, recent research has explored the integration of additional non-textual conditions—such as camera motion, depth maps, and human pose—to extend pretrained video generation models and enable more controllable video synthesis. These approaches aim to enhance the flexibility and practical applicability of AIGC-driven video generation systems. In this survey, we provide a systematic review of controllable video generation, covering both theoretical foundations and recent advances in the field. We begin by introducing the key concepts and commonly used open-source video generation models. We then focus on control mechanisms in video diffusion models, analyzing how different types of conditions can be incorporated into the denoising process to guide generation. Finally, we categorize existing methods based on the types of control signals they leverage, including single-condition generation, multi-condition generation, and universal controllable generation. For a complete list of the literature on controllable video generation reviewed, please visit our curated repository at \url{https://github.com/mayuelala/Awesome-Controllable-Video-Generation}.}




\maketitle
\section{Introduction}

As interest in AI-generated content (AIGC) continues to grow, video generation, one of its key domains, has emerged as a prominent focus for both researchers and users alike. Modern video generation methods~\citep{ho2022imagen, ma2024followposeposeguidedtexttovideo, chen2023videocrafter1opendiffusionmodels, z1, z2, z3, z4, jiang2025personalized, chen2024echomimiclifelikeaudiodrivenportrait, zhang2025mind, feng2025dit4edit,feng2025follow,  geng2024motionprompting, zhu2025aether, zhu2024instantswap,long2025follow} typically leverage cutting-edge generative paradigms (\textit{e.g.}, diffusion~\citep{ho2020ddpm,lipman2022flow,wang2025multishotmaster} or autoregressive models~\citep{van2016conditional, z5, z6, z7, ramesh2021dalle, chang2022maskgit, z10, tian2024var}), combined with large-scale datasets~\citep{Bain21WebVid-10M, chen2024panda-70m, schuhmann2021laion-400m}, massive model parameters~\citep{kong2025hunyuanvideosystematicframeworklarge, huang2025step, z9, wan2025}, and advanced architectural frameworks~\citep{Peebles2022DiT}. We refer to these models as video generation foundation models, which have significantly advanced the quality of generated videos. The resulting outputs exhibit an unprecedented level of creativity. Despite their impressive generative capabilities, these models often remain constrained by their reliance on text-only conditioning, which limits the degree of control users can exert over the generated content. As a result, users frequently struggle to translate their creative ideas into precise video outputs, thereby diminishing the practical effectiveness of these models in real-world content creation scenarios.


To address this challenge, researchers have begun exploring ways to incorporate control signals beyond text, enabling more accurate and flexible guidance in the video generation process. For example, enabling users to modify camera trajectories or specify particular actions for characters in the video are emerging areas of interest. When fine-grained control over the generated content becomes possible, users are empowered with greater creative flexibility, thereby unlocking the full potential and practical value of video generation as a task.

\begin{figure}[!t]
  \centering
  \includegraphics[width=1\linewidth]{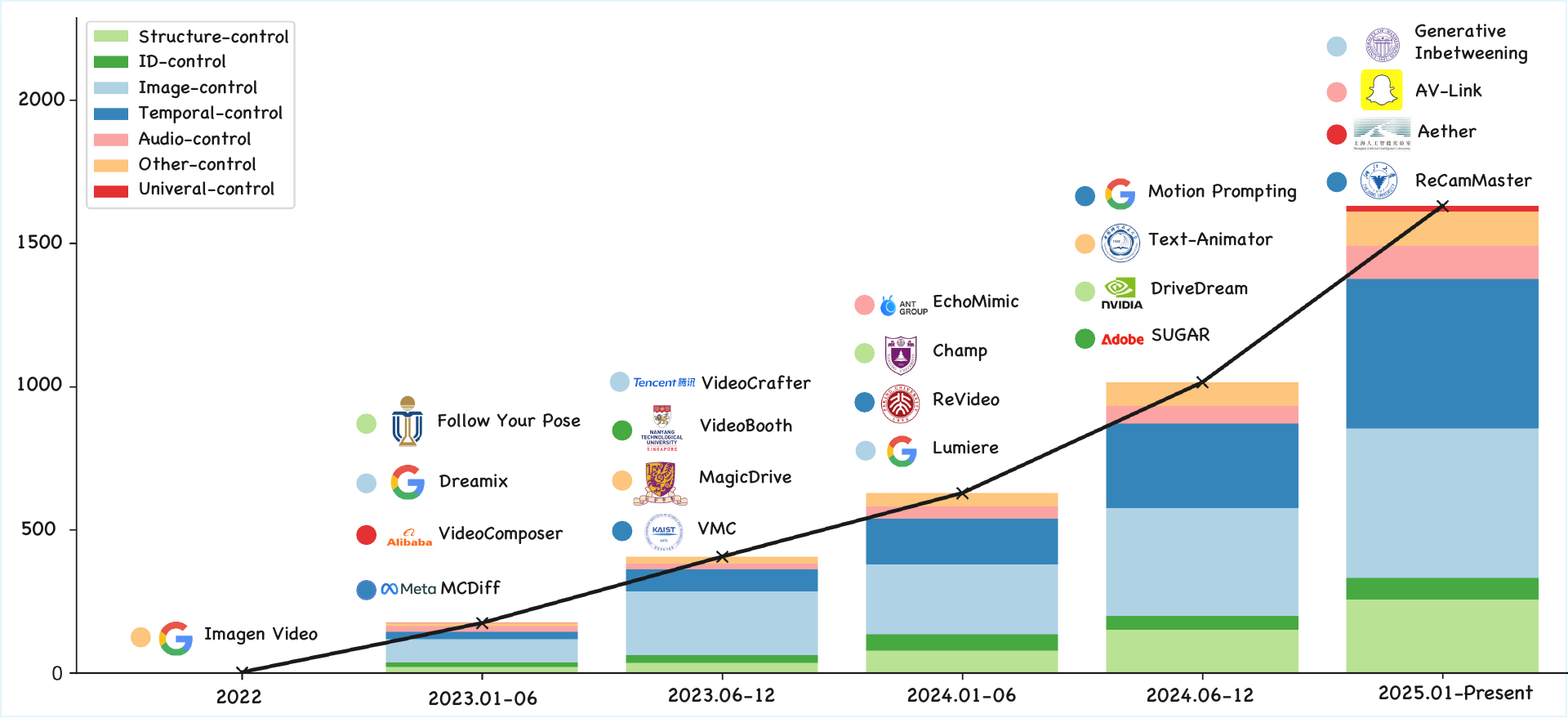}
  \caption{\textbf{The development trend of controllable video generation methods across seven representative task categories.} The line chart illustrates the rapid growth in the number of related works from 2022 to the present, with different categories distinguished by color. Representative works from each period are highlighted above the chart. For instance, VideoCrafter~\citep{chen2023videocrafter1opendiffusionmodels} and EchoMimic~\citep{chen2024echomimiclifelikeaudiodrivenportrait} have achieved 4.9k and 3.9k stars in Github, respectively. }
  \vspace{-20pt}
  \label{fig:paper_count}
\end{figure}

In this survey, we focus on the task of controllable video generation, including both its theoretical foundations and practical applications. Our goal is to \textit{provide a comprehensive overview of the latest research advances and to shed light on the development trajectory of this rapidly evolving field}. Specifically, we start by providing a brief overview of the background and core concepts of video generative models, providing their theoretical basis. This analysis clarifies the core principles of earlier research, fostering a deeper understanding of the field. Subsequently, the detailed reviews of previous studies are conducted to emphasize their unique contributions and distinctive features. Then we investigate the wide-ranging applications of these methods, highlighting their practical significance and influence across various contexts and related downstream tasks. Additionally, we deep discuss the limitations and future work about controllable video generation. In Fig.~\ref{fig:paper_count}, we present a line chart illustrating the number of controllable video generation studies utilizing various types of conditioning. As video foundation models have rapidly advanced, controllable video generation has also experienced significant growth.

Recent survey papers provide extensively overviews of AI-generated content (AIGC), covering various areas such as video generation based on Generative Adversarial Networks and Variational 
AutoEncoders~\citep{10.1145/3487891, yazdani2024comprehensive}, diffusion model theories and architectures~\citep{xing2024survey}, efficient video diffusion models~\citep{ulhaq2022efficient}, unified multi-modal video synthesis and understanding~\citep{zhou2024survey}, video editing~\citep{sun2024diffusion}, foundational video diffusion models~\citep{wang2025survey,yang2025videograin, melnik2024video}, and 4D generation applications~\citep{ren2025gen3c}. While these reviews offer valuable insights, many only provide a cursory examination of video generative models or predominantly concentrate on other modalities.  This is a significant gap in the literature regarding controllable video generation. Additionally, existing studies rarely address this topic in various control signals, \textit{e.g.}, depth, sketch, leaving a critical void in understanding the potential for integrating novel conditions into video generative models and their implications for advancing controllable video generation.

In summary, our contributions are as follows:

\begin{itemize}
    \item A well-structured taxonomy of controllable video generation methods is presented by classifying existing methods according to their input control signals, which facilitates understanding of existing methods and reveals core challenges in this field.

    \item We present the theoretical foundations of GAN-, VAE-, Flow-, DM-, and AR-based architectures, along with recent video generation models built upon them, providing a clearer understanding of their underlying mechanisms.

    \item Our survey introduces broad coverage of conditional generation approaches, structured around the proposed taxonomy, and emphasizes the defining traits and methodological innovations of each technique.
    
    \item We investigate the practical impact of conditional generation within video models, covering a range of generative scenarios that reflect its increasing significance in the AIGC landscape. In addition, we identify key shortcomings of existing techniques and propose potential avenues for further exploration.
    
\end{itemize}

The remainder of this paper is organized as follows: Sec.~\ref{sec:preliminaries} provides a concise overview of various generative paradigms. In Sec.~\ref{sec:fundation model}, we introduce representative video generation models, and presents a comprehensive taxonomy for controllable video generation. In Sec.~\ref{sec:methodologies}, we outline different control mechanisms, explain how novel conditions can be incorporated into video generation models, and summarize existing methods based on our proposed taxonomy. Sec.~\ref{sec:Application} highlights key application scenarios of controllable video generation. Lastly, in Sec.~\ref{sec:discussion and future}, we discuss several limitations of current research from both technical and practical perspectives, and propose promising directions for future work.

\section{Preliminaries}

\begin{figure}[th]
  \centering
  \includegraphics[width=0.95\linewidth]{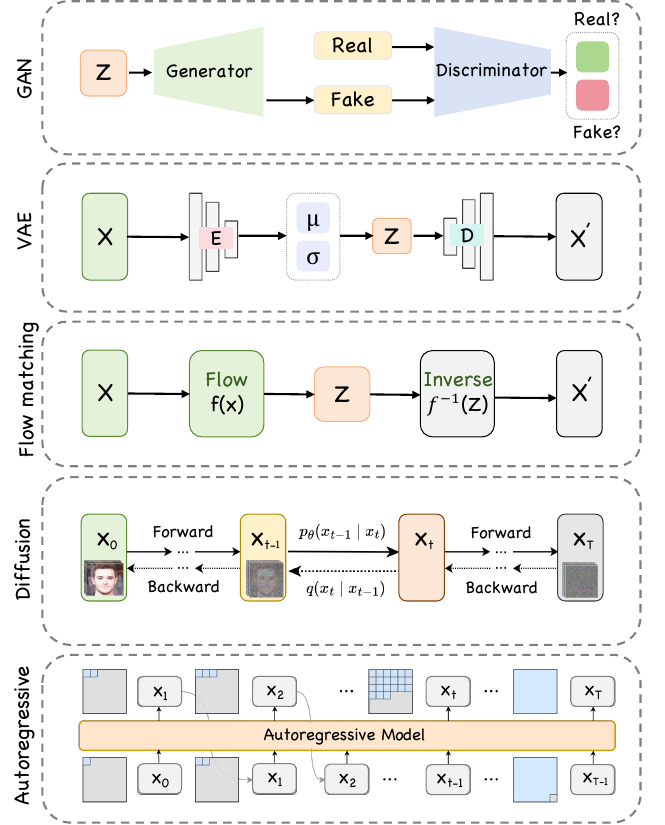}
  \caption{\textbf{Illustration of various generative methods.}
  We show the method of GAN, VAE, Flow Matching, Diffusion Models, and Autoregressive Models for video generation.}
  \label{fig:video_model}
\end{figure}

This section first explains the basic theories of generative models. In Fig.~\ref{fig:video_model}, we present an illustration of GAN, VAE, Flow-based Models, Diffusion Models, and Autoregressive Models, then we give the taxonomy of the controllable video generation tasks.

\label{sec:preliminaries}

\textbf{GANs and VAEs.}
Generative Adversarial Networks (GANs) and Variational Autoencoders (VAEs) are two classical models for generative modeling.
GAN~\citep{goodfellow2014generative} is based on a minimax game between a generator $G$ and a discriminator $D$. The generator tries to produce realistic samples from random noise $\mathbf{z} \sim p(\mathbf{z})$, while the discriminator attempts to distinguish between real data $\mathbf{x} \sim p(\mathbf{x})$ and generated samples $G(\mathbf{z})$. The objective is given by

\begin{equation*}
\begin{aligned}
\min_G \max_D \;\;
&\mathbb{E}_{\mathbf{x}}
\big[ \log D(\mathbf{x}) \big]
+ \mathbb{E}_{\mathbf{z}}
\big[ \log \big( 1 - D(G(\mathbf{z})) \big) \big].
\end{aligned}
\end{equation*}

VAE~\citep{kingma2013auto} introduces a variational posterior $q(\mathbf{z}|\mathbf{x})$ to approximate the true posterior $p(\mathbf{z}|\mathbf{x})$. The training objective is to maximize the Evidence Lower Bound (ELBO):


\begin{equation*}
\begin{aligned}
\mathcal{L}(\theta;\mathbf{x})
&= \mathbb{E}_{q(\mathbf{z}\mid \mathbf{x};\theta)}
\big[ \log p(\mathbf{x}\mid \mathbf{z};\theta) \big] \\
&\quad - D_{\mathrm{KL}}
\big( q(\mathbf{z}\mid \mathbf{x};\theta) \,\|\, p(\mathbf{z}) \big).
\end{aligned}
\end{equation*}

Both models form the foundation for developments in generative modeling, including diffusion and flow-based models.

\textbf{Diffusion Models.}
In recent years, diffusion models have emerged as a powerful series of generative models, offering high-quality sample generation. A typical model is the Denoising Diffusion Probabilistic Model (DDPM)~\citep{ho2020ddpm}, which add Gaussian noise to pure data and learn to reverse this process at sample generation, following the Markov Chain.

\begin{itemize}
    \item \textbf{Forward Process.} In the forward process, DDPM converts clean data from the previous data distribution to noise by gradually adding random Gaussian noise:

\vspace{-10pt}
\begin{equation*}
\begin{alignedat}{2}
q(x_1,\dots,x_T \mid x_0) &= \prod_{t=1}^{T} q(x_t \mid x_{t-1}), \\
q(x_t \mid x_{t-1})       &= \mathcal{N}\!\left(
x_t;\, \sqrt{1-\beta_t}\, x_{t-1},\, \beta_t \mathbf{I}
\right).
\end{alignedat}
\end{equation*}

    where the noise schedule $\beta_t$ is designed to increase monotonically with $t$, ensuring noise is smoothly added into the clean data.  
    \item \textbf{Reverse Process.} In the reverse process, the models aim to learn the ability to restore data from noise through denoising $x_t$. At each step, the denoising operation is formulated as
\vspace{-5pt}
\begin{equation*}
\begin{alignedat}{2}
p_\theta(x_{0:T})
&= p(x_T) \prod_{t=1}^{T} p_\theta(x_{t-1} \mid x_t), \\
p_\theta(x_{t-1} \mid x_t)
&= \mathcal{N}\!\left(
x_{t-1};\, \mu_\theta(x_t, t),\, \Sigma_\theta(x_t, t)
\right).
\end{alignedat}
\end{equation*}
\end{itemize} 
where the mean $\mu_{\theta}(x_t,t)$ and the variance $\Sigma_{\theta}(x_t,t)$ are parameterized by neural networks trained to predict the denoising transformation from $x_t$ to $x_{t-1}$. 
To simplify the reverse process, the diffusion models often reparameterize the mean $\mu_\theta(x_t, t)$ using a noise prediction model $\epsilon_\theta(x_t, t)$, which directly estimates the noise added at each timestep instead of recovering the clean data:
$$
\mu_\theta\left(x_t, t\right)=\frac{1}{\sqrt{\alpha_t}}\left(x_t-\frac{\beta_t}{\sqrt{1-\bar{\alpha}_t}} \epsilon_\theta\left(x_t, t\right)\right).
$$

\textbf{Flow-based Models.}
Flow-based models are a class of powerful generative models, with the core idea being to learn an accurate, invertible transformation that converts a simple base distribution into a complex target data distribution. This transformation process is typically composed of a series of invertible functions, referred to as ``flows".

Traditional flow-based models~\citep{dinh2014nice,chen2024follow, dinh2016density,kingma2018glow, zhang2025easycontrol, song2022cliptexture, song2023clipvg, song2022clipfont, chen2025transanimate} rely on discrete sequences of invertible transformations. In contrast, Continuous Normalizing Flows (CNFs)~\citep{chen2018neural} generalize this framework by modeling the transformation as a continuous-time flow governed by an Ordinary Differential Equation (ODE):
\vspace{-2pt}
$$
\begin{aligned} 
\frac{d}{d t} \phi_t(x) =v_t\left(\phi_t(x)\right), \quad 
\phi_0(x) =x, 
\end{aligned}
$$
where $\phi_t$ is the flow map, describing the transformation of data from $x_0$ to $x_t$, and $v_t$ is a time-dependent vector field, typically parameterized by neural networks, that defines the direction and magnitude of change.

To improve the efficiency of training CNFs, a modern generative approach called \textit{Flow Matching} (FM)~\citep{lipman2022flow} has been proposed. It offers a new paradigm for learning complex data distribution by directly learning the vector field that transfers samples from the base to the target distribution.

To learn the vector field $v_t$, Flow Matching introduces a novel supervision signal based on sample pairs $(x_0, x_1)$ drawn from the base and target distributions, respectively. It defines a straight-line interpolation between the two:
\vspace{-2pt}
$$
x_t=(1-t) x_0+t x_1,  v_t^{\mathrm{gt}}\left(x_t\right)=x_1-x_0,
$$
The model is then trained by minimizing the squared error between the predicted and ground-truth velocity:
\vspace{-2pt}
$$
\mathcal{L}_{\mathrm{FM}}=\mathbb{E}_{x_0 \sim p_0, x_1 \sim p_1, t \sim u[0,1]}\left[\left\|v_t\left(x_t\right)-\left(x_1-x_0\right)\right\|^2\right].
$$
This loss encourages the vector field to match the direction and magnitude of the straight-line transport from $x_0$ to $x_1$.

\textbf{Autoregressive Models.}
Autoregressive (AR)~\citep{ramesh2021dalle, Wu2021NÜWAVS, ding2021CogView,yu2022Parti, gafni2022makeascene, chang2022maskgit, tian2024var} Models generate data by modeling the conditional distribution of each element given the previous ones. Formally, given a sequence $\mathbf{x} = (x_1, x_2, \dots, x_T)$, the joint probability can be factorized as
\vspace{-2pt}
$$
P(\mathbf{x}) = \prod_{t=1}^T P(x_{t+1}|x_{\leq t}).
$$

In the early stage, AR models like PixelCNN~\citep{van2016conditional} generate pixel row by row, to capture local dependencies. While, due to their inherently sequential generation process, suffering from efficiency, they were gradually replaced by diffusion models. Recently, with the growth of video generation and the strong temporal correlations between frames, AR models have regained attention. Many works adopt hybrid designs, integrating autoregressive priors with diffusion backbones.

\section{Method}

\label{sec: method}

\begin{figure}[th]
    \centering
\includegraphics[width=0.5\textwidth]{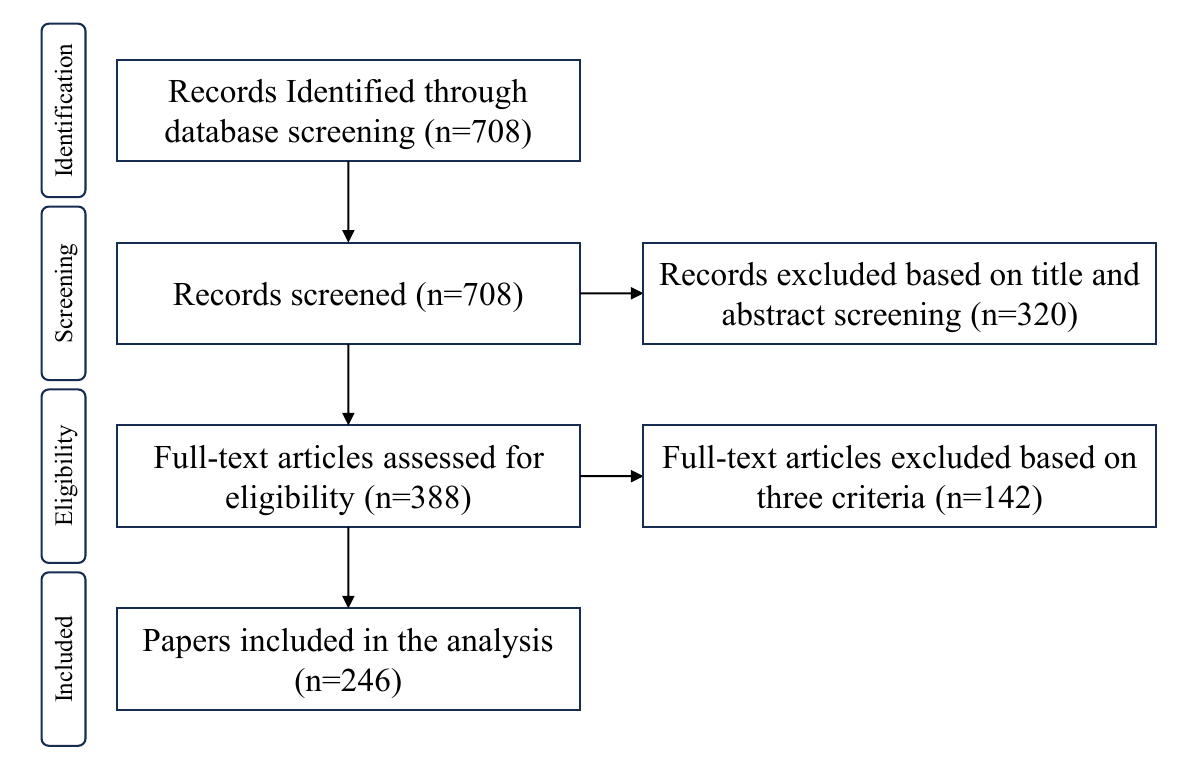}
\caption{\textbf{PRISMA~\citep{moher2009preferred} flow diagram of our literature review process.} We performed a four-phase screening covering identification, screening, eligibility assessment, and final inclusion, as detailed in Sec.~\ref{sec: method}. The number of papers retained at each stage is denoted by N.}
    \label{fig:prisma}
  \vspace{-10pt}
\end{figure}

In this section, we describe the methodology for our literature collection and screening process. We focused on papers published within the past five years (2020–2025), as this period marks the rapid evolution of controllable video generation techniques driven by diffusion models. As illustrated in Fig.~\ref{fig:prisma}, we followed the PRISMA~\citep{moher2009preferred} guideline with a four-phase procedure to select and review relevant works. To align with our research scope, we established the following three inclusion criteria: \textbf{C1. Controllable Video Generation:} The paper must present methods, frameworks, or systems that enable fine-grained control over video synthesis, beyond standard text-to-video generation. Acceptable control signals include but are not limited to: pose, depth, sketches, bounding boxes, motion trajectories, camera parameters, audio, identity, or multi-modal conditions. \textbf{C2. Based on Generative Foundation Models:} The paper should build upon widely recognized generative paradigms, such as DMs, ARs, GANs, flows, or hybrid architectures. \textbf{C3. Relevant Contributions:} We also include papers that introduce datasets, evaluation benchmarks, systematic analyses, or user-centric tools that facilitate controllable video generation, even if they do not propose new generative models.

\textbf{Phase 1: Identification.}
To identify influential works in controllable video generation, we selected a series of publication venues based on their impact in computer vision and graphics: Conference on Computer Vision and Pattern Recognition (CVPR), International Conference on Computer Vision (ICCV), European Conference on Computer Vision (ECCV), Neural Information Processing Systems (NeurIPS), International Conference on Learning Representations (ICLR), International Conference on Machine Learning (ICML), ACM Multimedia (MM), Winter Conference on Applications of Computer Vision (WACV), International Conference on Acoustics, Speech, and Signal Processing (ICASSP), International Conference on Robotics and Automation (ICRA), SIGGRAPH/SIGGRAPH Asia, ACM Transactions on Graphics (TOG), IEEE Transactions on Pattern Analysis and Machine Intelligence (TPAMI), Transactions on Visualization and Computer Graphics (TVCG), Transactions on Multimedia (TMM), and Transactions on Machine Learning Research (TMLR). We also considered high-impact preprints on arXiv to capture the most recent advances. We constructed a set of search queries using keywords reflecting the two core aspects of our survey: (1) \emph{controllability} and (2) \emph{video generation}. Example queries used in ACM Digital Library and arXiv search are:

\textbf{Title/Abstract: ((controllable video) OR (controlled video generation) OR (conditional video synthesis) OR (video control) OR (motion control) OR (camera control) OR (pose-guided video) OR (audio-driven video) OR (text-to-video) OR (image-to-video)) AND (generative model OR diffusion OR VAE OR GAN OR AR)}.


We included papers published between January 2020 and October 2025, written in English. The initial search yielded n=708 candidate papers.

\textbf{Phase 2: Screening.}
We screened the titles and abstracts of the 708 papers against criteria \textbf{C1–C3}. Papers that clearly did not address controllable video generation (\textit{e.g.}, focusing solely on video recognition, compression, or non-generative tasks) were excluded. Additionally, papers that introduced generative models without any mechanism for external control were filtered out. After this phase, 388 papers remained for full-text review. To ensure consistency, two authors independently screened a random subset of 100 papers; the inter-rater agreement measured by Cohen's Kappa was 0.87.

\textbf{Phase 3: Eligibility.}
We assessed the full text of the 388 papers for eligibility. Exclusion at this stage was based on: (a) failure to meet both \textbf{C1} and \textbf{C2} upon detailed reading, or (b) limited relevance to the core theme of controllable video generation despite meeting \textbf{C1} or \textbf{C2} superficially. For instance, papers that only briefly mentioned controllability as a future direction without substantial technical contribution were excluded. This phase removed 142 papers, leaving 246 papers for final inclusion.

\begin{figure*}
    \centering
    \includegraphics[width=\linewidth]{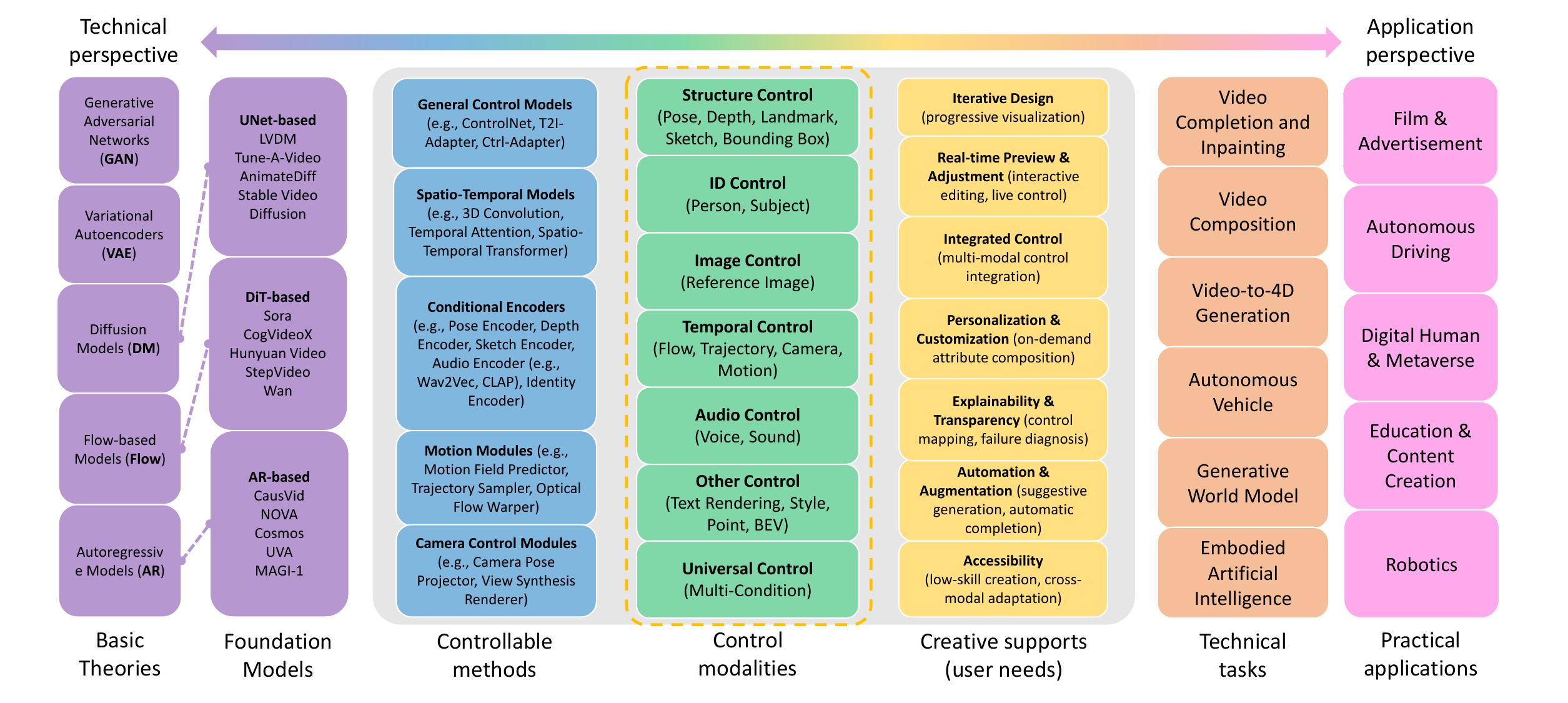}
    \caption{\textbf{An overall framework for controllable video generation.} The seven control modalities form the primary classification axis. The framework connects basic theories (Sec.~\ref{sec:preliminaries}) and video generation foundation models (Sec.~\ref{sec:fundation model}) with controllable methods and creative supports as key insights (Sec.~\ref{sec:methodologies}), and maps them to downstream technical tasks and practical applications (Sec.~\ref{sec:Application}).}
    \label{fig:video_taxonomy}
\end{figure*}

\textbf{Phase 4: Taxonomy.}
\label{sec: coding}
To provide a comprehensive and structured overview of the field, we propose a multi-layered framework that captures the entire ecosystem of controllable video generation, spanning from technical foundations to practical applications. As illustrated in Fig.~\ref{fig:video_taxonomy}, our framework encompasses six interconnected dimensions: \textbf{Technical Foundations:} The generative paradigms that form the basis of modern video synthesis, including GANs, VAEs, DMs, Flows, ARs, and corresponding foundation models. \textbf{Controllable Modules:} The specialized components designed to inject external conditions into pre-trained generative models, such as ControlNet, adapters, encoders, and other modules. \textbf{Control Modalities:} The types of input signals used to guide video generation, categorized into seven distinct classes based on their semantic and structural characteristics. \textbf{Creative Support:} The user-centric capabilities that empower creators through iterative design, real-time preview and adjustment, integrated control, etc. \textbf{Technical Tasks:} The core computer vision tasks enabled by controllable generation, including video completion and inpainting, video composition, video-to-4D generation, etc. \textbf{Practical Applications:} The real-world domains where these technologies demonstrate significant impact, such as film production, autonomous driving, and digital humans.

Within this comprehensive framework, we adopt \textbf{control modality} as our primary taxonomic dimension for three reasons. First, the type of control signal fundamentally determines the architectural design choices, conditioning mechanisms, and technical challenges in video generation pipelines. Second, existing research efforts are predominantly organized around specific control types such as pose-guided animation or audio-driven synthesis, making this classification naturally align with the current literature structure. Third, control modalities serve as the critical bridge between user intent (expressed through various input forms) and the generated visual content. As shown in Fig.~\ref{fig:video_taxonomy}, we systematically categorize controllable video generation methods into seven modality classes: (1) \textit{Structure Control} (\textit{e.g.}, pose, depth, sketch, bounding boxes), (2) \textit{ID Control} (person or subject identity), (3) \textit{Image Control} (reference images), (4) \textit{Temporal Control} (flow, trajectory, camera, motion), (5) \textit{Audio Control} (voice and sound), (6) \textit{Other Control} (text rendering, style, points, BEV), and (7) \textit{Universal Control} (multi-condition integration).

The task of controllable generation with text-to-video diffusion models represents a multifaceted and complex field. From the condition perspective, we divide this task into seven sub-stasks . Most works study how to generate video under specific conditions, \textit{e.g.} pose-guided generation, and depth-guided generation. To reveal the underlying mechanisms and characteristics of these approaches, we further categorize them according to their condition types. The primary challenge in this feild is enabling pretrained T2V diffusion models to handle novel types of conditions and generate videos that are not only aligned with the provided textual prompts but also exhibit high visual quality and strong temporal consistency. Moreover, recent methods inverstagte how to generate videos using multiple conditions, such as given a reference image, sparse trajectory, and motion brush. The main challenge in these tasks is the effective integration of multiple, potentially interacting, conditions, necessitating the capability to express several conditions simultaneously and consistently throughout the generated video sequence. 


\begin{table*}[!th]
\centering
\caption{\textbf{An overview of notable video generative models.} Resolution: The frame count, width and height of generated video. $f$: downsampling factor of autoencoder from pixel space to latent space. $^{\dag}$: The method uses one-shot data from the Dataset.}
\resizebox{\textwidth}{!}{%
\begin{tabular}{@{}llccccccc@{}}
\toprule
\textbf{Model} & \textbf{Venue} & \textbf{Param.} & \textbf{Resolution} & \textbf{\textit{f}} & \textbf{Text Encoder} & \textbf{Training Dataset} & \textbf{Open Source} &\textbf{Latency}\\ \midrule
\multicolumn{8}{l}{\textit{UNet-Based Video Diffusion Models}} \\
\midrule
LVDM~\citep{he2022lvdm} & arXiv 2022 & 1.2B & $\text{Multi-Frame(16\&1024)}\times256^2$ & 4, 8x8 & CLIP ViT-L/14 & WebVid-2M~\citep{Bain21WebVid-10M} & \ding{51} & 114 \\ 

Tune-A-Video~\citep{wu2023tune} & ICCV 2023 & 983M & $24\times512^2$ & 1, 8x8 & CLIP ViT-L/14 & DAVIS$^{\dag}$
~\citep{pont-tuset2017DAVIS}& \ding{51} & 769\\ 

AnimateDiff~\citep{guo2023animatediff} & ICLR 2024 & 1.4B & $16\times256^2$ & 1, 8x8 & CLIP-ViT-L/14 & WebVid-10M~\citep{Bain21WebVid-10M} & \ding{51} & 122\\ 

Stable Video Diffusion~\citep{blattmann2023stablevideodiffusionscaling} & arXiv 2023 & 1.5B & $14\times576\times1024$ & 1, 8x8 & CLIP-ViT-H/14  & Internal Dataset & \ding{51} & 98\\ 

\midrule
\multicolumn{8}{l}{\textit{DiT-Based Video Diffusion Models}} \\ 
\midrule


CogVideoX~\citep{yang2025cogvideoxtexttovideodiffusionmodels}& ICLR 2025 & 2B\&5B & $49\times480\times720$ & 4, 8x8 & T5 & Internal Dataset & \ding{51} & 180\\ 
HunyuanVideo~\citep{kong2025hunyuanvideosystematicframeworklarge} & arXiv 2024 & 13B & $129\times720\times1280$ & 4, 8x8 & Hunyuan-Large & Internal Dataset & \ding{51} & 3448\\ 



StepVideo~\citep{huang2025step} & arXiv 2025 & 30B & $204\times544\times992$ & 8, 16x16 & Hunyuan-CLIP, Step-LLM & Internal Dataset & \ding{51} & 743 \\ 
Wan~\citep{wan2025} & arXiv 2025 & 1.3B\&14B & $81\times720\times1280$ & 4, 8x8& umT5 & Internal Dataset & \ding{51} & 1864\\ 
%
\midrule
\multicolumn{8}{l}{\textit{ Video Autoregressive Models}} \\
\midrule
CausVid~\citep{yin2025slowbidirectionalfastautoregressive} & CVPR 2025 & 1.4B & $120\times352\times640$ & 4, 8x8 & umT5 & Internal Dataset & \ding{51} & 14\\ 
NOVA~\citep{deng2025autoregressivevideogenerationvector} & ICLR 2025 &  0.3B\&0.6B\&1.4B  & $33\times768\times480$ & 4, 8x8 & Phi-2 & Internal Dataset & \ding{51} & 354\\ 
Cosmos~\citep{nvidia2025cosmosworldfoundationmodel} & arXiv 2025 & 4B\&12B & $121\times704\times1280$ & 4, 8x8 & T5-XXL & Internal Dataset & \ding{51} & 119\\ 
UVA~\citep{li2025unifiedvideoactionmodel} & RSS 2025 & 0.5B & $16\times720\times1280$ & 1, 16x16 & CLIP-ViT-B/32 & Libero10~\citep{liu2023libero}, PushT~\citep{chi2024diffusionpolicy}, UMI~\citep{chi2024UMI}, Human Video~\citep{clark2025actionfreereasoningpolicygeneralization} & \ding{51} & <1 \\ 


MAGI-1~\citep{ai2025magi1autoregressivevideogeneration} & arXiv 2025 & 4.5B\&24B & $24\times720\times1280$ & 4, 8x8 & T5 & Internal Dataset & \ding{51} &1740\\ 

InfinityStar~\citep{liu2025infinitystar} & arXiv 2025 & 8B & $81\times720\times1280$ & 4, 8x8 & T5 & Internal Dataset & \ding{51} & 58 \\

\bottomrule
\end{tabular}%
}
\vspace{-10pt}
\label{tab:base_models}
\end{table*}

\section{Video Generation Foundation Models}
\label{sec:fundation model}
Video generation foundation models have recently attracted substantial attention due to their remarkable ability to generate high-fidelity videos. These models are typically categorized into two major paradigms: diffusion-based models and autoregressive (AR) models. Among them, diffusion-based methods have shown impressive performance by modeling data distributions through an iterative denoising process, effectively capturing complex spatial-temporal dependencies. We introduce the typical frameworks among them (see in Table~\ref{tab:base_models}). Diffusion-based video generation models can be further divided into two architectural branches: those built on UNet-based frameworks and those leveraging the more recent DiT (Diffusion Transformer) architecture. We provide them in supplementary material.

\section{Controllable Video Generation Models with Various Conditions}
\label{sec:methodologies}

\begin{figure}[ht]
    \centering
\includegraphics[width=0.5\textwidth]{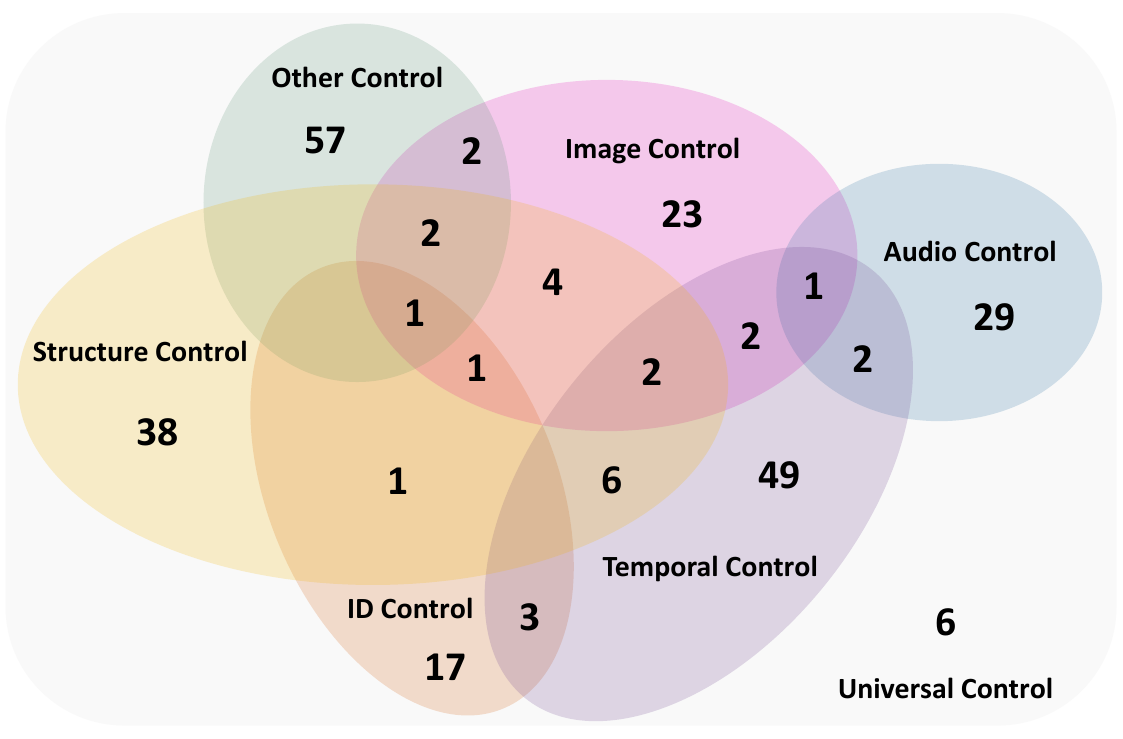}
\caption{\textbf{Venn diagram of controllable video generation literature categorized by control modalities.} The diagram illustrates the distribution of surveyed works across seven control categories, including Structure, Image, Audio, Temporal, ID, Universal, and Other controls. The overlap regions highlight methods that support multiple control signals simultaneously. The counts in each region indicate the number of papers falling into the corresponding category intersections, reflecting current research trends and the degree of multimodal controllability explored in existing video generation models. }
\vspace{-15pt} 
\label{fig:venn}
\end{figure}

\begin{figure*}
  \centering
  \includegraphics[width=1.0\linewidth]{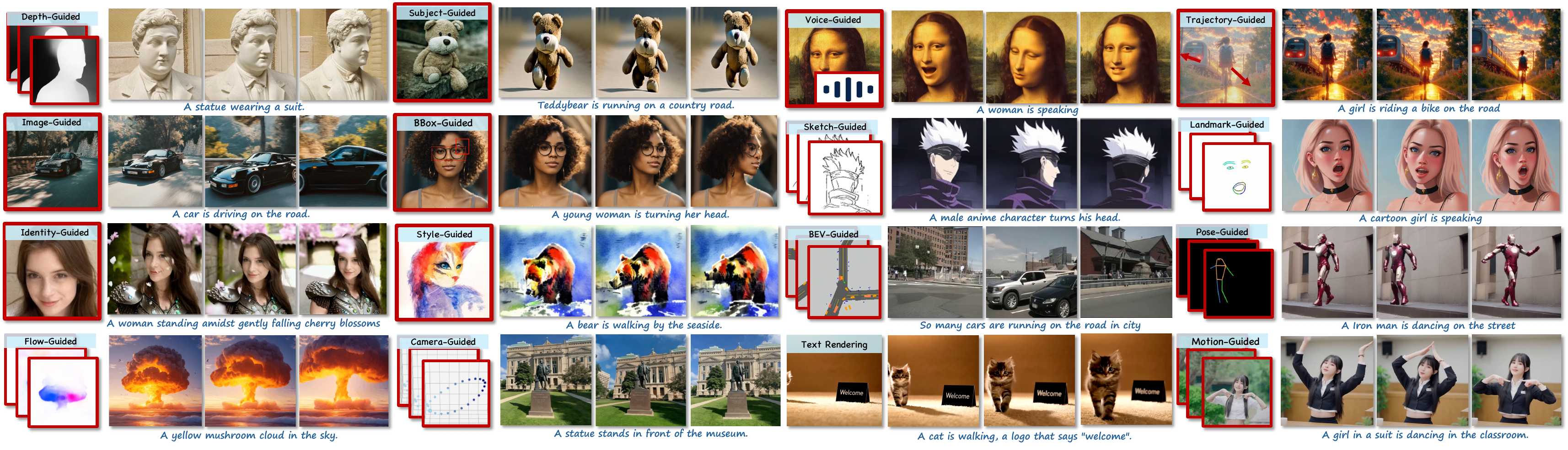}
  \caption{\textbf{Visual illustration of controllable video generation.} We show the cases of controllable video generation with specific control. The conditions are marked in \textbf{red} and the prompts are marked in \textbf{blue}.}
  \label{fig:visual_illustration}
\end{figure*}

This section forms the core of our survey, delving into the diverse methodologies developed for controllable video generation. As shown in Fig.~\ref{fig:visual_illustration} and Table.~\ref{tab:methods}, we present various visual illustrations and representative works of controllable video generation. These approaches are categorized on the basis of the primary nature of the control signal used to guide the synthesis process. Specifically, we will explore methods focusing on structure control, ID control, image control, temporal control, audio control, other control, and universal control. For each category, we will review seminal and recent works, highlighting their foundational techniques, architectural innovations, and the specific challenges.

\subsection{Structure Control}
\label{sec:method_structure_control}
Structure control in video generation refers to the ability to dictate the spatial layout, conformation of articulated objects (like humans or animals), and the geometric properties of the scene. This form of control is paramount for producing videos that are not only visually coherent but also adhere to plausible physical configurations and user-specified arrangements. By conditioning the generation process on structural cues, models can synthesize complex scenes with greater fidelity and semantic correctness.

\subsubsection{Pose-Guided Generation}
\label{sec:method_pose_guided}
Pose-guided video generation specifically aims to animate subjects, predominantly humans, according to a sequence of predefined poses. This is a critical task for applications such as virtual avatar animation, character generation for interactive entertainment, human motion synthesis from sparse inputs, and fashion video synthesis~\citep{karras2023dreamposefashionimagetovideosynthesis, jin2024alignment, wang2025diffdecompose, gong2025relationadapter, ma2024magicstickcontrollablevideoediting, ma2025followcreation, wang2024cove}. The core challenge lies in generating temporally coherent video frames where the subject's appearance is preserved while accurately following the target pose sequence $P = {p_1, p_2, ..., p_T}$, where each $p_t$ represents the pose in frame $t$.

The input pose information $p_t$ is typically represented as 2D skeletal keypoints (\textit{e.g.}, COCO format), 3D skeletal coordinates, or more dense representations like DensePose~\citep{guler2018densepose}, which maps image pixels to 3D surface coordinates of the human body. These pose sequences can be extracted from existing videos using off-the-shelf pose estimators (\textit{e.g.}, OpenPose~\citep{cao2019openpose}), derived from motion capture (MoCap) data, or generated algorithmically. The choice of pose representation influences the granularity of control and the complexity of the generation task.

Numerous approaches have been proposed to achieve pose-guided generation. Early methods~\citep{yang2018poseguidedhumanvideo} often rely on GANs, using pose information to condition the generator, for instance, by rendering pose stick figures and feeding them as input alongside a latent code. More recent work leverages the power of diffusion models. For example, MagicAnimate~\citep{xu2023magicanimatetemporallyconsistenthuman} and Animate Anyone~\citep{hu2024animateanyoneconsistentcontrollable} and its successor Animate Anyone 2~\citep{hu2025animate2highfidelitycharacter} utilize diffusion models conditioned on pose sequences and a reference image to generate temporally consistent animations. Techniques like Champ~\citep{zhu2024champ} and MimicMotion~\citep{zhang2024mimicmotionhighqualityhumanmotion} focus on high-fidelity human motion synthesis. DirectorLLM~\citep{song2024directorllmhumancentricvideogeneration} employs large language models to direct human-centric video generation based on pose and textual descriptions. Follow Your Pose~\citep{ma2024followposeposeguidedtexttovideo} introduces a two-stage training strategy to create high-quality character videos. ControlNet-based approaches~\citep{zhang2023addingconditionalcontroltexttoimage} have also been adapted, where pose maps serve as direct spatial conditions for pre-trained text-to-image diffusion models, which are then fine-tuned or augmented with temporal modules for video tasks~\citep{peng2024controlnext, wang2024discodisentangledcontrolrealistic}. Disentangling pose from appearance is a key strategy, often achieved by dedicated appearance encoders and pose encoders, with mechanisms like feature warping or attention to align appearance features with the target pose~\citep{li2025disposedisentanglingposeguidance}. Temporal consistency is often enforced through temporal attention mechanisms across frames or by incorporating temporal smoothness losses.

\begin{table*}[!t]

\caption{\textbf{Representative works of controllable generation with video generative model.}
}
\vspace{-10pt}
\center
\small
\setlength{\tabcolsep}{5pt}
\resizebox{\linewidth}{!}{
\begin{tabular}{c | l | c | c | c | c }
\toprule
\textbf{Type} &
\textbf{Method} & \textbf{Venue} & \textbf{Model} & \textbf{Condition} & \textbf{Training Dataset} \\
\toprule

\multirow{5}{*}{Structure Control} 

&Follow-Your-Pose~\cite{ma2024followposeposeguidedtexttovideo} & AAAI 2024 & UNet & Pose, Landmark, Text & LAION-400M~\cite{schuhmann2021laion-400m}, HDVILA~\cite{xue2021hdvila}\\
&Make-Your-Video~\cite{xing2024make} & TVCG 2024& UNet & Depth, Text & WebVid-10M~\cite{Bain21WebVid-10M}\\
&ToonCrafter~\cite{10.1145/3687761} & TOG 2024 & UNet & sketch, Image, Text & Self-Construction Datasets\\
&DriveDreamer~\cite{zhao2025drivedreamer} & AAAI 2024 & UNet & BBox, Image, Text & nuScenes~\cite{caesar2019nuScenes} \\
&EchoMimic~\cite{meng2024echomimicv2strikingsimplifiedsemibody} & AAAI 2024 & UNet & Audio, Landmarks, Image & HDTF~\cite{zhang2021HDTF}, CelebV-HQ~\cite{zhu2022CelebV-HQ}\\
\midrule 

\multirow{4}{*}{ID Control}
&VideoBooth~\cite{jiang2023videobooth} & CVPR 2023 &  UNet & Object, Text & WebVid-10M~\cite{Bain21WebVid-10M} \\
&Vlogger~\cite{zhuang2024vlogger} & CVPR 2024 &  UNet & Object, Text & WebVid-10M~\cite{Bain21WebVid-10M}, LAION-400M~\cite{schuhmann2021laion-400m} \\

&Phantom~\cite{liu2025phantom} & ICCV 2025 &  DiT & Subject, Text & Panda-70M~\cite{chen2024panda-70m}, Subject200k~\cite{tan2024subject200k}, OmniGen~\cite{xiao2024OmniGen}\\


\midrule


\multirow{5}{*}{Image Control}
&VideoCrafter1~\cite{chen2023videocrafter1opendiffusionmodels} & arXiv 2023 & UNet & Image, Text &LAION-COCO-600M~\cite{LAION-COCO}, WebVid-10M~\cite{Bain21WebVid-10M}  \\

&Lumiere~\cite{10.1145/3680528.3687614} & SIGGRAPH-Asia 2024 & UNet & Image, Mask, Style, Text &  Self-Construction Datasets\\

&ConsistI2V~\cite{ren2024consisti2venhancingvisualconsistency} & TMLR 2024 & UNet & Image, Text& WebVid-10M~\cite{Bain21WebVid-10M} \\

&Follow-Your-Click~\cite{ma2024followyourclickopendomainregionalimage} & AAAI 2025 & UNet & Click, Image, Text& WebVid-10M~\cite{Bain21WebVid-10M}\\

&NOVA~\cite{deng2025autoregressivevideogenerationvector} & ICLR 2025 & Autoregressive & Image, Text & Panda-70M~\cite{chen2024panda-70m}, Pexels~\cite{Pexels} \\

\midrule

\multirow{4}{*}{Temporal Control}

&Motion-I2V~\cite{10.1145/3641519.3657497} & SIGGRAPH 2024 & UNet & Trajectory, Image, Text & WebVid-10M~\cite{Bain21WebVid-10M} \\
&MotionBooth~\cite{wu2024motionbooth} &NeurIPS 2024& UNet & Motion, Camera, BBox, Image, Text&Panda-70M~\cite{chen2024panda-70m}\\
&ViewCrafter~\cite{wu2024motionbooth} &TPAMI 2025& UNet & Camera, Image &RealEstate10K~\cite{tinghui2018RealEstate10K}, DL3DV~\cite{ling2023DL3DV}\\
&Direct-A-Video~\cite{yang2024directavideo} & SIGGRAPH 2024& UNet & Camera, Text & MovieShot~\cite{rao2020MovieShot} \\

\midrule

\multirow{4}{*}{Audio Control}

&DAA2V~\cite{yariv2024diverse} &AAAI 2023& UNet & Audio, Image &VGGSound~\cite{Chen20VGGSound}, Landscape~\cite{lee2022Landscape}, AudioSet-Drums~\cite{gemmeke2017AudioSet} \\
&MOFA-Video~\cite{niu2024mofavideo} & ECCV 2024 & UNet&Trajectory, Audio, Flow, Image & WebVid-10M~\cite{Bain21WebVid-10M}\\
&EMO~\cite{tian2024emoemoteportraitalive} & ECCV 2024 & UNet & Audio, Image & HDTF~\cite{zhang2021HDTF}, VFHQ~\cite{xie2022VFHQ}, CelebV-HQ~\cite{zhu2022CelebV-HQ}\\
&MotionCraft~\cite{bian2025motioncraft} & AAAI 2025 & DiT & Music, Speech, Text & HumanML3D~\cite{Guo_2022_CVPR_HumanML3D}, BEAT2~\cite{liu2023BEAT2}, FineDance~\cite{li2023finedance}\\

\midrule

\multirow{4}{*}{Other Control}

&Panacea~\cite{wen2024panacea} & CVPR 2023 & UNet & BEV, Text & nuScenes~\cite{caesar2019nuScenes}\\
&StyleMaster~\cite{ye2024stylemaster} & CVPR 2025& DiT & Style, Video, Text & Self-Construction Datasets \\
&UniVST~\cite{song2024univst} &TPAMI 2025 & DiT & Style, Mask, Video & None\\
&GS-DiT~\cite{bian2025gs} & CVPR 2025 & DiT & Point, Video, Text & WebVid-10M~\cite{Bain21WebVid-10M}\\

\midrule

\multirow{3}{*}{Universal Control}

&VideoComposer~\cite{wang2023videocomposer} & NeurIPS 2023 & UNet & Universal Conditions & WebVid-10M~\cite{Bain21WebVid-10M}, LAION-400M~\cite{schuhmann2021laion-400m}\\

&FullDiT~\cite{ju2025fulldit} & ICCV 2025 & DiT & Universal Conditions & MiraData~\cite{ju2024miradata}, RealEstate10K~\cite{tinghui2018RealEstate10K}, ConceptMaster~\cite{huang2025conceptmaster}, Panda-70M~\cite{chen2024panda-70m}\\

&VACE~\cite{jiang2025vace} & ICCV 2025 & DiT & Universal Conditions & Self-Construction Datasets\\
\bottomrule 
\end{tabular}
}
\vspace{-10pt}
\label{tab:methods}
\end{table*}

\par Despite significant progress, pose-guided generation still faces challenges. Handling severe self-occlusions, where parts of the subject are hidden due to the pose, requires robust inpainting capabilities. Maintaining fine-grained texture details and consistent identity across large pose variations remains difficult, especially when appearance must be reliably transferred from a reference image without explicit identity control. Ensuring natural and smooth transitions between poses, avoiding jittery or robotic movements, is crucial for realism. Furthermore, generalization to unseen poses, body shapes, or complex clothing items not well-represented in training data can be problematic, often requiring extensive and diverse datasets like those proposed by FCVG~\citep{zhu2024generativeinbetweeningframewiseconditionsdriven}, which aims for broader character controllability.

\subsubsection{Depth-Guided Generation}
\label{sec:method_depth_guided}
    Depth-guided video generation utilizes depth maps as conditioning signals to control the 3D structure of the synthesized scene, object layout, and relative distances of elements from the camera. This form of control is crucial for generating videos with a strong sense of three-dimensionality, realistic object interactions, and consistent scene geometry across frames.

    The primary input for depth-guided control is a sequence of depth maps $D = {d_1, d_2, ..., d_T}$, where each $d_t \in \mathbb{R}^{H \times W}$ provides per-pixel depth information. These depth maps can be relative or absolute, estimated from RGB videos using monocular depth estimation models, captured using specialized sensors (\textit{e.g.}, LiDAR, ToF cameras), or rendered from 3D models. Depth conditioning helps in maintaining geometric consistency, enabling plausible occlusions, and can be used to simulate effects like depth-of-field.
    
    Several works have explored depth-guided video synthesis. ControlVideo~\citep{zhang2023controlvideo} and Control-A-Video~\citep{chen2023control} demonstrate how various control signals, including depth, could be integrated into large pre-trained text-to-image diffusion models to guide video generation. SparseCtrl~\citep{guo2024sparsectrl} focuses on using sparse depth controls. DreamDance~\citep{pang2024dreamdance} leverages depth for dynamic 3D scene generation. ControlNeXt~\citep{peng2024controlnext} and Champ~\citep{zhu2024champ} also support depth guidance among other control types. Depth information is typically encoded and injected into the diffusion model\'s U-Net architecture, often concatenated with noisy latents or fed into cross-attention layers. Some methods may also use depth-based loss functions to further encourage geometric fidelity, for instance, by comparing the depth map of a generated frame with the target depth map.

    Challenges in depth-guided generation include the difficulty of obtaining accurate, dense, and temporally consistent depth maps, especially from monocular RGB videos. Errors or noise in the input depth maps can propagate to the generated video, leading to distorted geometry. Ensuring that the generated RGB frames are consistent with the conditioning depth information is non-trivial. Dynamic scenes with rapidly changing depths or complex non-rigid object deformations pose further difficulties. Moreover, incorporating 3D awareness often increases computational complexity. Approaches like Make-Your-Video~\citep{xing2024make} and Moonshot~\citep{zhang2024moonshot} aim to improve fidelity and consistency in such scenarios.


\subsubsection{Landmark-Guided Generation}
\label{sec:method_landmark_control}

Landmark-guided control focuses on using keypoints or landmarks (\textit{e.g.}, facial features, body joints, or other salient points) to guide video synthesis. These landmarks serve as spatial constraints that define the motion or deformation of specific regions, enabling fine-grained control over facial expressions, gestures, or body poses. This approach is particularly valuable for applications like portrait animation, emotion-driven synthesis, and pose-guided video generation, where subtle movements and precise alignment with landmarks are critical for achieving natural and expressive results.

Landmark control signals can be provided as static or dynamic landmark sequences extracted from images or videos, or as high-level descriptions that are converted into landmark trajectories. For instance, TASTE-Rob~\citep{zhao2025tasterobadvancingvideogeneration} uses temporal landmark sequences to guide video generation. Follow-Your-Pose~\citep{ma2024followposeposeguidedtexttovideo} and Follow-Your-Emoji~\citep{ma2024followyouremojifinecontrollableexpressivefreestyle} allow pose-based and emoji-driven landmark control, respectively, offering users intuitive ways to define movements and expressions.

Recent landmark-guided methods leverage landmarks to condition generative models, ensuring spatial alignment and temporal coherence. HunyuanPortrait~\citep{xu2025hunyuanportraitimplicitconditioncontrol} introduces implicit conditioning mechanisms to generate highly realistic and controllable portrait animations. Takin-ADA~\citep{lin2024takinadaemotioncontrollableaudiodriven} and EchoMimic~\citep{chen2024echomimiclifelikeaudiodrivenportrait} specialize in audio-driven portrait animation, aligning facial landmarks with speech dynamics. EchoMimicV2~\citep{meng2024echomimicv2strikingsimplifiedsemibody} extends this by simplifying body and facial control for efficient synthesis. LivePortrait~\citep{guo2025liveportraitefficientportraitanimation} and IPT2V~\citep{yuan2024identitypreserving} emphasize identity preservation while animating portraits, ensuring that synthesized videos remain faithful to the input identity. 

Landmark-guided methods face significant challenges in maintaining accurate spatial alignment, especially when landmarks are sparse, noisy, or incomplete. Ensuring smooth and natural transitions between frames while adhering to landmark constraints is critical for avoiding artifacts and achieving realistic motion. Capturing subtle expressions or complex dynamics, such as emotion or speech-driven facial movements, can be difficult without introducing distortions or losing fidelity.

\subsubsection{Sketch-Guided Generation}
Sketch-guided video generation utilizes single sketch frames or combines them with reference visual images to improve the performance of traditional text-to-video diffusion models. Sketch input becomes particularly valuable in applications such as quick character animation, where users can sketch the key poses or actions to guide the video generation process, and scene composition, where simple sketches can define object placement and movement in a visual story. This approach is especially beneficial for colorization applications like storytelling, allowing creators to quickly visualize and refine their ideas before detailed production.

In previous works, works like SketchBetween~\citep{Loftsdottir_2022} introduce a VAE-based framework for the sketch-based video generation application. Recently, the ControlNet structure~\citep{zhang2023addingconditionalcontroltexttoimage} has been applied to sketch-guided video generation, but directly using it does not ensure the frame-to-frame consistency essential for coherent video production. As the existing control method commonly relies on dense temporal maps, SparseCtrl~\citep{guo2024sparsectrl} proposed an auxiliary encoder like ControlNet, which utilizes the sparse condition maps to generate the target video. Similarly, CTRL-Adapter~\citep{lin2024ctrladapterefficientversatileframework} designed a control adapter that supports various inputs, including image, depth, and sketch, facilitating the performance of controllable video generation applications. OpenSoraPlan~\citep{lin2024opensoraplanopensourcelarge} proposed a powerful video generation framework with a condition encoder.

Numerous works also focus on exploring sketch-guided video generation applications, such as animation, video colorization, and many other tasks. TaleCrafter~\citep{gong2023talecrafter}, a visual storytelling system, designs a controllable text-to-image component that generates target images based on input conditions and converts them into video through a subsequent Image-to-Video (I2V) module. ToonCrafter~\citep{10.1145/3687761} proposes a cartoon video interpolation method, which designs a sketch encoder to provide efficient drawing tools. AniDoc~\citep{meng2025anidocanimationcreationeasier} and LVCD~\citep{Huang_2024} also focus on the sketch-guided animation application. For general video generation, methods like MakePixelsDance~\citep{zeng2023makepixelsdancehighdynamic}, VideoLCM~\citep{wang2023videolcmvideolatentconsistency}, and TF-T2V~\citep{Wang_2024_CVPR}, these methods not only improve the quality and the controllability of the video but also optimize the consistency for the long-time video generation task. Besides, MagicStick~\citep{ma2024magicstickcontrollablevideoediting}, SketchVideo~\citep{Liu_2025_CVPR}, MotionClone~\citep{ling2024motionclonetrainingfreemotioncloning}, and CameraCtrl~\citep{he2024cameractrl} utilize the sketch-guided controllable component for video editing and camera motion video generation.

\begin{table*}[t!]
\centering
\vspace{-10pt}
\caption{\textbf{Quantitative comparison of Landmark-guided video generation on $512 \times 512$ test images.} LMD multiplied by \( 10^{-3} \), AED multiplied by \( 10^{-2} \), APD multiplied by \( 10^{-3} \) and Identity Similarity multiplied by \( 10^{-1} \). ${\uparrow}$ indicates higher is better. ${\downarrow}$ indicates lower is better. The best results are in bold.}
\resizebox{1.0\linewidth}{!}
{
\begin{tabular}{lcccccc|cccccc}
\toprule
\multirow{3}{*}{\textbf{Method}}  & \multicolumn{6}{c}{\textbf{Self Reenactment}} & \multicolumn{6}{c}{\textbf{Cross Reenactment}} 
\\ 
\cmidrule(lr){2-7}  \cmidrule(lr){8-13}
 & \multirow{2}{*}{\textbf{LMD}\ $\downarrow$} & \multirow{2}{*}{\textbf{FID-VID}\ $\downarrow$} & \multirow{2}{*}{\textbf{FVD}\ $\downarrow$} & \multirow{2}{*}{\textbf{PSNR}\ $\uparrow$} & \multirow{2}{*}{\textbf{SSIM}\ $\uparrow$} & \multirow{2}{*}{\textbf{LPIPS}\ $\downarrow$} & \multirow{2}{*}{\textbf{AED}\ $\downarrow$} & \multirow{2}{*}{\textbf{APD}\ $\downarrow$} & \textbf{Identity}\ $\uparrow$  & \textbf{Facial}\ $\uparrow$ & \textbf{Video}\ $\uparrow$ & \textbf{Temporal} \ $\uparrow$
 \\ 
 & & & & & &  &   &  & \textbf{Similarity} & \textbf{Movements} & \textbf{Quality} & \textbf{Smoothness} \\
\midrule
LivePortrait~\citep{guo2025liveportraitefficientportraitanimation} & 9.14 & 82.71 & 483.38 & 31.41 & 0.72 & 0.22 & 26.83 & 26.97 & 8.71 & 3.00 & 4.11 & 4.06 \\
AniPortrait~\citep{wei2024aniportraitaudiodrivensynthesisphotorealistic} & 6.67 & 81.90 & 430.24 & 30.54 & 0.67 & 0.27 & 22.42 & 21.32 & 7.95 & 3.83 & 4.00 & 3.81 \\
FollowYE~\citep{ma2024followyouremojifinecontrollableexpressivefreestyle} & 5.63 & 77.15 & 417.53 & 30.84 & 0.68 & 0.25 & 20.06 & 20.95 & 8.18 & 4.11 & 4.23 & 3.33 \\
X-Potrait~\citep{xie2024x} & 6.23 & 82.93 & 416.41 & 30.81 & 0.71 & 0.19 & 20.66 & 20.39 & 8.03 & 3.88 & 3.73 & 3.49 \\
HunyuanPortrait~\citep{xu2025hunyuanportraitimplicitconditioncontrol} & \textbf{2.02} & \textbf{75.81} & \textbf{333.48} & \textbf{32.98} & \textbf{0.81} & \textbf{0.11} & \textbf{19.45} & \textbf{19.20} & \textbf{8.87} & \textbf{4.55} & \textbf{4.69} & \textbf{4.61}  \\ 
\bottomrule
\end{tabular}
}
\vspace{-5pt}
\label{tab:metrics}
\end{table*}

\subsubsection{BBox-Guided Generation}
BBox-guided video generation directs the synthesis of video content by using bounding boxes as precise conditional inputs. This guidance can range from a static layout defining object placement in a single frame to a sequence of moving boxes dictating an object's trajectory over time. The technique is prominently explored in autonomous driving video models, where bounding boxes are used to control the overall layout of the environment, as well as the precise position and movement of vehicles within the scene.

To inject bbox information into generative models, researchers have developed various strategies, which can generally be categorized into two main types: cross-attention-based injection and additive encoder-based injection. MagicDrive~\citep{gao2023magicdrive}, aiming for fine-grained control over 3D geometry, pioneers the separate handling of different layout types. 
Delphi~\citep{ma2024unleashing} also injects layout embeddings via cross-attention. Another early work, DriveDreamer~\citep{zhao2025drivedreamer}, seeking to build a real-world driving model, fuses the representations of 3D bounding boxes ($B_i \in \mathbb{R}^{N \times N_B \times 16}$ ) with visual features using a gated self-attention mechanism. 
To tackle the challenge of learning box-object correlations from purely visual cues, Boximator~\citep{wang2024boximator} and Panacea~\citep{wen2024panacea} introduce the new self-attention layer into the existing U-Net blocks, which processes control tokens derived from Fourier-encoded box coordinates and object IDs.
Ctrl-V~\citep{luo2025ctrl} also utilizes a ControlNet architecture but introduces a unique two-stage pipeline. 
It first employs a diffusion-based BBox Generator to predict a complete video of the bounding box trajectories. 

Moreover, some methods explore different injection methods. LLM-Grounded Video Diffusion (LVD)~\citep{lian2023llm}  takes a novel approach by leveraging a Large Language Model (LLM) to generate Dynamic Scene Layouts from text prompts. 
Also operating without training, TrailBlazer~\citep{ma2024trailblazer} directly edits spatial and temporal attention maps during the initial denoising steps to guide subjects along key-framed bounding box trajectories. 
Recently, architectures based on the DiT have gained attention for their excellent scalability. For instance, DIVE~\citep{jiang2024dive}, aiming to generate multi-view consistent videos, integrates a ControlNet-Transformer into the DiT architecture to process road information and uses a joint cross-attention mechanism to fuse scene and instance layouts. 

\subsubsection{Summary}
In video generation, structural control acts as spatial cues, providing the foundational framework for dynamic visuals. Its core function is to precisely translate various abstract spatial requirements—from precise joint poses to casual user sketches—into natural and plausible visual movements.    Early GAN-based frameworks~\citep{yang2018poseguidedhumanvideo} often conditioned the generation process by directly rendering structural skeletons together with latent features. And this paradigm was subsequently dominated by {UNet-based architectures} which employs Adapter interfaces (\textit{e.g.}, \textit{ControlNet}~\citep{zhang2023addingconditionalcontroltexttoimage})to graft spatial constraints onto main network. Currently, the research focus has clearly shifted towards diffusion Transformers(\textit{e.g.}, \textit{MagicDrive-DiT}~\citep{gao2024magicdrivedit})due to the scaling limits of CNNs.

This shift is from the user's need for finer control granularity and temporal coherence, with trends leaning toward integrated approaches that enable combining multiple structural signals (\textit{e.g.}, pose with depth in DreamDance~\citep{pang2024dreamdance}). Beyond interaction, profound technical challenges persist in physical awareness and generalization. Current methods often prioritize visual alignment over kinematic plausibility, leading to physically impossible deformations, and remain limited to specific domains (\textit{e.g.}, humans). Consequently, future improvement is needed for  {interaction simplicity} and physics-aware world modeling. The ultimate goal is to bridge the gap between abstract control and concrete realization, ensuring outcomes are not only physically consistent but also faithfully aligned with the user's creative intent.

\begin{table*}[t]
\centering
\caption{\textbf{Quantitative comparison of landmark-guided video generation on TED-talks Dataset.} L1 multiplied by \( 10^{-4} \). ${\uparrow}$ indicates higher is better. ${\downarrow}$ indicates lower is better. The best and secondary results are in bold and underline, respectively. MRAA* and TPSMM* proposed these methods using real videos as driving signals. Therefore, they are not involved in quantitative comparison.}


\resizebox{0.8\textwidth}{!}{
\begin{tabular}{lccccccc}
\toprule
\textbf{Method} & \textbf{FID} $\downarrow$ & \textbf{SSIM} $\uparrow$ & \textbf{PSNR} $\uparrow$  & \textbf{LPIPS} $\downarrow$ & \textbf{L1} $\downarrow$ & \textbf{FID-VID} $\downarrow$ & \textbf{FVD} $\downarrow$ \\
\midrule

MRAA*\citep{siarohin2021motion} & \color{gray}25.76 & \color{gray}0.807 & \color{gray}32.62 & \color{gray}0.216 & \color{gray}0.36 & \color{gray}12.20 & \color{gray}126.28 \\
TPSMM*\citep{zhao2022thin} & \color{gray}24.69 & \color{gray}0.819 & \color{gray}32.92 & \color{gray}0.205 & \color{gray}0.35 & \color{gray}9.04 & \color{gray}87.76 \\

MRAA~\citep{siarohin2021motion} & 50.36 & 0.762 & \underline{31.90} & 0.266 & 0.50 & 82.79 & 493.02 \\
TPSMM~\citep{zhao2022thin} & 23.71 & {0.771} & {\textbf{32.30}} & 0.252 & \underline{0.49} & 32.12 & 260.67 \\
DisCo~\citep{wang2024discodisentangledcontrolrealistic} & 75.48 & 0.575 & 27.99 & 0.309 & 1.21 & 66.18 & 393.04 \\
MagicAnimate~\citep{xu2023magicanimatetemporallyconsistenthuman} & 41.58 & 0.529 & 28.28 & 0.310 & 1.73 & 33.61 & 223.54 \\
MagicPose~\citep{chang2024magicposerealistichumanposes} & \underline{23.39} & 0.723 & 30.08 & \underline{0.236} & 0.81 & \underline{27.53} & \underline{214.23} \\
FollowYourPosev2~\citep{xue2024follow} & {\textbf{18.21}} & {\textbf{0.779}} & 30.88 & {\textbf{0.198}} & {\textbf{0.46}} & {\textbf{10.24}} & {\textbf{81.73}} \\
\bottomrule
\end{tabular}}
\end{table*}

\subsection{ID Control}
ID control in video generation refers to the ability to preserve and manipulate the visual identity of specific entities within the whole video generation, such as humans, animals, or objects. This form of control is essential for ensuring temporal consistency in appearance, enabling personalized or instance-specific content generation. By conditioning on reference images or identity embeddings, models can maintain subject integrity across varying poses, motions, and viewpoints.

\label{sec:personalization}
\subsubsection{Person-Guided Generation}
Person-guided video generation aims to synthesize videos of a specific individual while maintaining high-fidelity identity consistency throughout the video. It enables personalized animation where a character—typically provided as one or more reference images—is animated according to user-defined prompts or motion signals. This form of control is essential for digital avatars, identity-preserving storytelling, and interactive applications, where both visual likeness and behavioral alignment must be achieved.

The primary input for person-guided control includes one or more ID images of the target person. These can be facial portraits or full Face and body synthesis. A driving signal—such as a text prompt, pose sequence, or audio—is also provided to guide motion. The ID images capture static visual identity, while the driving signal controls dynamics. Some methods, like ID-Animator\citep{he2024idanimator}, use a single image, while others, like Movie Weaver\citep{liang2025movieweaver} rely on a set of images covering both face and body.

Representative diffusion-based methods explore various architectures and identity-injection strategies. 
ID-Animator\citep{he2024idanimator} integrates a ViT-based face adapter via cross-attention, while Movie Weaver\citep{liang2025movieweaver} leverages a DiT backbone with anchor-concept prompts for tuning-free generation. 
PersonalVideo\citep{li2025personalvideo} and Vlogger~\citep{zhuang2024vlogger} applie a reward-based objective to a frozen LDM for identity-motion alignment. ConsisID~\citep{yuan2024identitypreserving} and Magic-Me~\citep{ma2024magicme} introduce frequency-aware and dynamic attention mechanisms.


\subsubsection{Subject-Guided Generation}
Subject-guided generation~\citep{zhang2024ssr,jiang2023videobooth,song2025layertracer,song2025makeanything} refers to the process of generating video content where the primary focus is on guiding the generation with specific subjects, such as particular characters, objects, or scenes, to ensure that the generated video adheres to user-defined subject constraints. This method enables the customization of content by explicitly controlling which subjects appear, how they interact, and in what context, offering more precise control over the generated content.
Initial research often concentrated on single-subject customization, as exemplified by VideoBooth~\citep{jiang2023videobooth}, which utilizes image prompts for enhanced content control, and DreamVideo~\citep{DreamVideo}, which decouples subject learning from motion learning. However, the field has rapidly progressed to address the more complex challenge of multi-subject video generation. Studies such as DisenStudio~\citep{chen2024disenstudio}, CustomVideo~\citep{CustomVideo}, and 
MovieWeaver~\citep{liang2025movieweaver}
are at the forefront of addressing key difficulties in multi-subject scenarios. These challenges include ensuring the correct co-occurrence and temporally consistent appearance of multiple subjects, preserving their visual identities, avoiding attribute binding issues, and accurately assigning actions. Such approaches typically extend pretrained video diffusion models by incorporating several key innovations: novel fine-tuning strategies (\textit{e.g.}, VideoDreamer's~\citep{chen2023videodreamer} Disen-Mix Finetuning, DisenStudio's~\citep{chen2024disenstudio} motion-preserved disentangled fine-tuning); advancements in attention mechanisms (\textit{e.g.}, CustomVideo's~\citep{CustomVideo} attention control with object segmentation); and the development of specialized datasets. 

To further enhance the quality, controllability, and usability of customized videos, researchers are exploring various innovative approaches. In pursuit of model efficiency and generalization, zero-shot or tuning-free methods have emerged, exemplified by SUGAR~\citep{SUGAR}, which leverages large-scale synthetic datasets.
ConsisID~\citep{yuan2024identitypreserving} offers a tuning-free method for identity preservation based on frequency decomposition. Still-Moving~\citep{Still-Moving} uniquely facilitates the customization of video models using solely image data through the training of spatial adapters. Another key research area is fine-grained control over motion and concepts. For instance, MotionBooth~\citep{wu2024motionbooth} focuses on precise object and camera movement control. CustomTTT~\citep{CustomTTT} employs test-time training to effectively combine multiple individually trained concepts (\textit{e.g.}, appearance and motion LoRAs), while CustomCrafter~\citep{CustomCrafter} aims to preserve the model's intrinsic motion generation and concept composition capabilities during subject learning.

\subsubsection{Summary} In this area, ID serves as a pivotal modality, utilizing conditional encoders and general control models (\textit{e.g.}, Adapters) to achieve the core goal of personalized customization. While UNet remains the dominant backbone (\textit{e.g.}, PersonalVideo~\citep{li2025personalvideo}), there is a strategic shift toward DiT (\textit{e.g.}, Movie Weaver~\citep{liang2025movieweaver}) to leverage global attention for improved temporal consistency.

Driven by the user demand for efficiency, the technical roadmap is transitioning from slow fine-tuning to Tuning-free approaches (\textit{e.g.}, ConsisID~\citep{yuan2024identitypreserving}), which directly aligns with the users' demand for real-time. However, current limitations in multi-subject generation reveal a gap in the unified system of creative supports, underscoring the future need for advanced disentanglement and composition to resolve conflicts between different controllable methods (\textit{e.g.}, conditional encoders and motion modules) without performance degradation.

\subsection{Image Control}
\label{sec:image}

The image-guided video generation aims to generate a video from a given reference image. This form of control is crucial for producing videos that are not only visually consistent with the reference but also maintain coherence across temporal frames. By conditioning the generation process on reference image features, models can generate video content with enhanced visual alignment, stylistic fidelity, and semantic relevance to the source image. 
Approaches such as TRIP~\citep{Zhang_2024_CVPR}, 
OmniTokenizer~\citep{wang2024omnitokenizerjointimagevideotokenizer}, Emu Video~\citep{girdhar2024emuvideofactorizingtexttovideo}, 
and I2V-Adapter~\citep{10.1145/3641519.3657407} focus on addressing the insufficient consistency between the given image and the generated video, constructing a unified, high-quality video generation framework. To further construct a comprehensive video generation model, STIV~\citep{lin2024stivscalabletextimage} and Hunyuan Video~\citep{kong2025hunyuanvideosystematicframeworklarge} are based on the Diffusion Transformer (DiT) framework with the image condition, which perform better in both T2V and I2V tasks.\par Another challenge is to keep the alignment between the reference and the given prompt. Specifically, at the start of the generated video, the image always plays a more crucial role for the details, style, and the object location; however, in the later stage, the effects of the reference image can easily be weakened by the prompt. To solve this problem, DreamVideo~\citep{10887583} introduces an Image Retention module, maintaining both information from the input image and prompt. AID~\citep{xing2024aidadaptingimage2videodiffusion} incorporates an MLLM (Multimodal Large Language Model) and DQFormer (Dual Query Transformer) to predict future frames based on the given key frame. DynamiCrafter~\citep{10.1007/978-3-031-72952-2_23} utilizes a dual-stream image injection paradigm to adapt various applications (\textit{e.g.}, animation, storytelling). Techniques like Cond image leak~\citep{zhao2024identifyingsolvingconditionalimage} and EDG~\citep{tian2025extrapolatingdecouplingimagetovideogeneration} aim to combat the issue of limited motion degrees and the unexpected motion with the prompt in the generated video. However, for the multi-object scenarios and the long-term video generation, the accuracy and consistency of the motion are also a challenge. Research like Through-The-Mask~\citep{yariv2025throughthemaskmaskbasedmotiontrajectories}, Lumiere~\citep{10.1145/3680528.3687614}, TI2V-Zero~\citep{ni2024ti2vzerozeroshotimageconditioning}, and EasyAnimate~\citep{xu2024easyanimatehighperformancelongvideo} attempt to utilize the powerful basic model (\textit{e.g.}, DiT) or improve the computation paradigm (\textit{e.g.}, compute all frames once and employ a special inversion strategy) while achieving a superior inference efficiency. 
\par Despite those remarkable achievements, image condition can also be easily combined with various conditions such as BBox, motion, and depth. Motion-I2V~\citep{10.1145/3641519.3657497} first deduces the potential motion from the reference image, and then uses both predicted motion and image to generate high-quality video. Decoupled I2V~\citep{shen2023decouplecontentmotionconditional} disentangles the motion vector and the image to obtain a memory-efficient and consistent video generation method. 
Similarly, MotionStone~\citep{shi2024motionstonedecoupledmotionintensity} also models the motion by the motion head and injects it into the Diffusion Transformer. Besides, MoVideo~\citep{liang2024movideomotionawarevideogeneration} utilizes the motion and keyframe with the depth and optical flow for video generation. ST-I2V~\citep{10.1145/3581783.3611897} cropped several areas from the reference image, using them to maintain the semantic information of the video. 
\par Differently, PhysGen~\citep{liu2024physgenrigidbodyphysicsgroundedimagetovideo} consideres the physical
properties like mass or elasticity, and external factors such as forces and environmental conditions. By giving input force and images as conditions, it can generate a video without a training process. HoloTime~\citep{zhou2025holotimetamingvideodiffusion} introduces a 360-degree 4D scene generation framework to reconstruct high-quality 4D scene video. ConsistI2V~\citep{ren2024consisti2venhancingvisualconsistency} concatenates the input image to the noise and optimizes the attention calculation, not only performing better in consistency, but also supports various condition inputs (\textit{e.g.}, layout, camera). To achieve more user-friendly video generation, video doodles~\citep{10.1145/3592413} and Follow-Your-Click~\citep{ma2024followyourclickopendomainregionalimage} allow users to insert hand-drawn animations and provide a click to select which area to move, benefiting from better interactivity.  Additionally, some work has explored using an image as the reference for other model architectures or tasks. NOVA~\citep{deng2025autoregressivevideogenerationvector} reformulates the video generation task as a non-quantized autoregressive modeling of temporal and spatial prediction, achieving a superior performance with fewer parameters. VideoPanda~\citep{xie2025videopandavideopanoramicdiffusion} also introduces a long video generation framework using auto-regression. Structure and Content-Guided Video Synthesis~\citep{Esser_2023_ICCV} 
extends the image condition to the video editing and video frame interpolation applications, respectively.

In summary, image control utilizes conditional encoders to map visual features into the backbone, fulfilling the user need for personalization and customization through high-fidelity visual alignment. In terms of the foundation models, the framework highlights a dynamic architectural ecosystem. Traditional UNet structures (\textit{e.g.}, I2V-Adapter~\citep{10.1145/3641519.3657407}) are increasingly being complemented by DiT (\textit{e.g.}, STIV~\citep{lin2024stivscalabletextimage}) and Autoregressive models (\textit{e.g.}, NOVA~\citep{deng2025autoregressivevideogenerationvector}), enabling superior scalability and temporal coherence for complex generation tasks.
However, persistent challenges in multi-object scenarios and prompt-image alignment suggest a continued need for better explainability to help users understand how models balance conflicting inputs during the generation process.

\subsection{Temporal Control}
\label{sec:temporal}
Temporal control in video generation refers to the ability to regulate the evolution of motion and timing across frames, ensuring that the generated content follows specific temporal dynamics. This form of control is crucial for synthesizing videos with coherent motion patterns, realistic pacing, and causally consistent frame transitions. By conditioning the generation process on temporal signals such as trajectories, action sequences, flow, or camera movements, models can produce videos that align with user-defined temporal intent.

\subsubsection{Flow-Guided Generation}
\label{sec:method_flow_control}

Flow-guided control utilizes optical flow or motion field representations to guide video synthesis, ensuring realistic temporal consistency and smooth object dynamics across frames. By incorporating motion cues, these methods provide finer control over movement and transitions, making them essential for coherent video generation tasks like animation and dynamic scene modeling.

Flow guidance can be specified through optical flow fields, motion trajectories, or displacement representations, either explicitly computed or implicitly learned. For example, Motion-I2V~\citep{10.1145/3641519.3657497} improves visual consistency by conditioning on motion trajectories, while MOFA-Video~\citep{niu2024mofavideo} animates static images using user-defined motion fields.

Recent approaches integrate flow-based guidance into generative architectures. I2VControl~\citep{feng2024i2vcontrol} ensures temporal coherence through disentangled flow representations, while MCDiff~\citep{chen2023motionconditioneddiffusionmodelcontrollable} employs motion-conditioned diffusion models for precise motion control. MOFA-Video~\citep{niu2024mofavideo} aligns motion fields for controllable animation, and Motion-I2V~\citep{10.1145/3641519.3657497} emphasizes flow-based trajectory conditioning to enhance consistency.

Challenges include maintaining motion consistency in complex scenes, avoiding artifacts from inaccurate flow, and efficiently integrating flow information without computational overhead. Techniques like MCDiff~\citep{chen2023motionconditioneddiffusionmodelcontrollable} and Motion-I2V~\citep{10.1145/3641519.3657497} address these issues but highlight the need for further advancements in flow-guided video generation.

\begin{table*}[htbp]
\centering
\renewcommand{\arraystretch}{1.0}
\caption{\textbf{Quantitative comparison of camera-guided video generation on 3D static and 4D dynamic scenes, using public and internet data.} We evaluate generation quality (FID, $\mathrm{CLIP}_{\mathrm{sim}}$ for static; FVD, $\mathrm{CLIP\text{-}V}_{\mathrm{sim}}$ for dynamic) and trajectory accuracy (ATE, RPE-T, RPE-R). All methods use official code with identical inputs. $\uparrow$ indicates higher is better, $\downarrow$ indicates lower is better. The best and secondary results are in bold and underline, respectively.}
\resizebox{\linewidth}{!}{
\begin{tabular}{l cc cc ccc ccc}
\toprule
\multirow{4}{*}{\textbf{Methods}} & \multicolumn{4}{c}{\textbf{Generation Quality}} & \multicolumn{6}{c}{\textbf{Trajectory Accuracy}} \\
\cmidrule(lr){2-5}\cmidrule(lr){6-11}
& \multicolumn{2}{c}{Static} & \multicolumn{2}{c}{Dynamic} & \multicolumn{3}{c}{Static} & \multicolumn{3}{c}{Dynamic} \\
\cmidrule(lr){2-3}\cmidrule(lr){4-5}\cmidrule(lr){6-8}\cmidrule(lr){9-11}
& FID $\downarrow$ & $\mathrm{CLIP}_{\mathrm{sim}} \uparrow$ 
& FVD $\downarrow$ & $\mathrm{CLIP\text{-}V}_{\mathrm{sim}} \uparrow$ 
& ATE $\downarrow$ & RPE-T $\downarrow$ & RPE-R $\downarrow$
& ATE $\downarrow$ & RPE-T $\downarrow$ & RPE-R $\downarrow$ \\
\midrule
{See3D \citep{ma2025you}} 
& 123.26 & \underline{0.941} &  -- & -- 
& 0.091 & \underline{0.089} & \underline{0.250} 
& -- & -- & -- \\
{ViewCrafter \citep{yu2024viewcrafter}} 
& 117.50 & 0.930 & -- & -- 
& 0.236 & 0.315 & 0.728 
& -- & -- & -- \\
{ViewExtrapolator \citep{liu2024novel}} 
& 125.50 & 0.930 & 108.48 & 0.913 
& 0.183 & 0.260 & 0.882 
& 1.040 & 1.208 & 4.750 \\
{TrajectoryAttention \citep{xiao2024trajectoryattention}} 
& 122.37 & 0.920 & 106.94 & 0.911 
& 0.159 & 0.238 & 0.532 
& 0.605 & 1.238 & 3.560 \\
{TrajectoryCrafter \citep{yu2025trajectorycrafter}} 
& \underline{111.49} & 0.910 & \underline{97.31} &\underline{0.923}
& \underline{0.090} & 0.152 & 0.267 
& \textbf{0.431} & 1.078 & 8.950 \\
{NVS\mbox{-}Solver \citep{you2024nvs}} 
& 118.64 & 0.937 & -- & -- 
& 0.224 & 0.268 & 1.056 
& -- & -- & -- \\
{WorldForge}\citep{song2025worldforge} 
& \textbf{96.08} & \textbf{0.948} & \textbf{93.17} & \textbf{0.938} 
& \textbf{0.077} & \textbf{0.086} & \textbf{0.221} 
& \underline{0.527} & \textbf{0.826} & \textbf{2.690} \\
\bottomrule
\end{tabular}
}
\label{tab:quality}
\end{table*}

\subsubsection{Trajectory-Guided Generation}
\label{sec:method_trajectory_control_merged}
Trajectory-guided video generation allows users to specify the motion paths of one or more objects or image regions across video frames. Applications include directing character movements along a specific route, animating inanimate objects with desired kinematics, or ensuring specific interactions between entities based on their spatial paths. 

Trajectory input can take various forms: a sequence of 2D or 3D coordinates $\{\mathbf{x}_t, \mathbf{y}_t, (\mathbf{z}_t)\}_{t=1}^T$ for a point of interest, bounding box tracks $\{\text{bbox}_t\}_{t=1}^T$ defining the extent and location of an object over time (as in Boximator~\citep{wang2024boximator}), user-sketched paths in an initial frame, or even textual descriptions of motion (\textit{e.g.}, ``move object A from left to right"). Drag-based interfaces, as explored in 
DragAnything~\citep{wu2024draganything} and DragEntity~\citep{wan2024dragentity}, 
allow interactive specification of start and end points for object parts.

Many recent methods leverage diffusion models for trajectory control. MotionBooth~\citep{wu2024motionbooth} and Motion-I2V~\citep{10.1145/3641519.3657497} focus on generating video from an image based on motion prompts or trajectories. TrailBlazer~\citep{ma2024trailblazer} and Direct-a-Video~\citep{yang2024directavideo} (which also handles camera control) enable explicit path following. Peekaboo~\citep{jain2024peekaboo} controls object appearance and disappearance along trajectories. MOFA-Video~\citep{niu2024mofavideo} focuses on fine-grained trajectory control. Trajectory information is often encoded as a sequence of coordinates or heatmaps and is used to guide the cross-attention layers or directly modify the latent representations in diffusion models. Warping fields or flow-based methods can also be guided by trajectories. MCDiff~\citep{chen2023motionconditioneddiffusionmodelcontrollable} introduced motion-controlled diffusion. More advanced concepts like Perception-as-Control~\citep{chen2025perceptionascontrol} and Motion Canvas~\citep{xing2025motioncanvas} provide sophisticated frameworks. LeviTor~\citep{wang2025levitor} and LayerAnimate~\citep{yang2025layeranimate} explore layered representations for controllable animation along trajectories.

Key challenges include generating motion that is not only accurate to the specified trajectory but also appears natural and physically plausible, avoiding jerky or sliding movements. Ensuring that the guided object interacts realistically with other scene elements and the background is difficult. Handling complex, intersecting, or occluded trajectories for multiple objects (as in Collaborative Video Diffusion~\citep{kuang2024collaborative}) requires sophisticated reasoning. Maintaining long-range temporal consistency of the object\'s appearance and the trajectory adherence over many frames is also crucial. 
Tora~\citep{zhang2025tora} and MagicMotion~\citep{li2025magicmotion} aim to address some of these complexities, with works like I2VControl~\citep{feng2024i2vcontrol} 
focusing on image-to-video with trajectory control.

\subsubsection{Camera-Guided Generation}
\label{sec:method_camera_control_merged}
Camera-guided video generation focuses on controlling the virtual camera's parameters to dictate the viewpoint, motion (\textit{e.g.}, pan, tilt, zoom, dolly, crane shots), and potentially intrinsic parameters (\textit{e.g.}, focal length) during video synthesis. This is essential for achieving specific cinematic styles, narrative perspectives, and dynamic visual effects, offering filmmakers and content creators fine-grained control over how a scene is presented.

Camera control signals can be provided as explicit sequences of camera poses (extrinsics: rotation $\mathbf{R}_t$ and translation $\mathbf{T}_t$, and intrinsics $\mathbf{K}_t$ if they vary), relative transformations between frames, or higher-level textual descriptions of camera movements (\textit{e.g.}, ``dolly zoom out from the character"). Motion Prompting~\citep{geng2024motionprompting} explores textual descriptions for camera and object motion.
Recent advancements heavily leverage diffusion models conditioned on camera parameters. CameraCtrl~\citep{he2024cameractrl} and its successor CameraCtrl II~\citep{he2025cameractrl2} provide explicit control over camera trajectories. Direct-a-Video~\citep{yang2024directavideo}, also mentioned for trajectory control, supports camera path guidance. ViewCrafter~\citep{yu2024viewcrafter} 
focuses on generating novel views based on camera inputs. 
CamCo~\citep{xu2024camco} and Trajectory Attention~\citep{xiao2024trajectoryattention} explore camera control in specific contexts. Uni3C~\citep{cao2025uni3cunifyingprecisely3denhanced} explores the joint control of human motion and camera trajectory. The camera parameters are typically used to define projection matrices that transform latent scene representations or guide the sampling of features to render the scene from the specified viewpoint. EgoSim~\citep{yu2025egosim} focuses on egocentric camera simulation. Approaches like I2VControl~\citep{feng2024i2vcontrol} , CamI2V~\citep{zheng2024cami2v}, and RealCam-I2V~\citep{li2025realcami2v} specialize in image-to-video generation with camera control. More advanced systems like 3DTrajMaster~\citep{fu2024trajmaster}, MotionFlow~\citep{zhang2025motionflow}, and CineMaster~\citep{wang2025cinemaster} aim for comprehensive 3D-aware camera and motion control. 

Challenges include maintaining 3D scene consistency and avoiding disocclusion artifacts or distorted geometry when the camera undergoes significant movement or viewpoint changes. Generating high-fidelity novel views that are consistent with previous frames is difficult. Ensuring smooth and natural camera transitions, free from jitter or abrupt changes, is crucial for cinematic quality. Providing intuitive user interfaces for specifying complex 3D camera paths and parameters remains an open area. 
ReCamMaster~\citep{bai2025recammaster} aims to improve the quality and control of camera movements. Techniques like VD3D~\citep{bahmani2025vd3d}, Aether~\citep{zhu2025aether}, and GenDoP~\citep{zhang2025gendop} explore complex 3D scene understanding and camera path generation. The latest work Voyager~\citep{huang2025voyagerlongrangeworldconsistentvideo} introduces a world caching scheme and smooth video sampling, and Follow-Your-Creation~\citep{ma2025followcreation} utilizes the temporal pack inference strategy to keep the long-term consistency, respectively.

\begin{table*}[t]
    \centering\small
    \caption{\textbf{Quantitative comparisons of camera-guided video generation on T2V setting across \textsc{RealEstate10K} and \textsc{DL3DV}.} Frame Consistency and Motion Strength are denoted as FC and MS. In addition. The rotation and translation error are denoted as RE and TE.}
    \setlength{\tabcolsep}{3pt}
    \resizebox{1\linewidth}{!}{
    \begin{tabular}{l|c|ccc|cc|c|ccc|cc}
        \toprule
        \multirow{2}{*}{\textbf{Method}} 
        & \multicolumn{6}{c|}{\textbf{RealEstate10K}} 
        & \multicolumn{6}{c}{\textbf{DL3DV}} \\
        \cmidrule(lr){2-7} \cmidrule(lr){8-13}
        & FVD $\downarrow$ & CLIPSIM $\uparrow$ & FC $\uparrow$ & MS $\uparrow$ & RE $\downarrow$ & TE $\downarrow$ 
        & FVD $\downarrow$ & CLIPSIM $\uparrow$ & FC $\uparrow$ & MS $\uparrow$ & RE $\downarrow$ & TE $\downarrow$ \\
        \midrule
        MotionCtrl \citep{wang2024motionctrl} & 506.9 & {0.2744} & 0.9471 & 0.9734 & 2.90 & 1.06 
                   & 746.7 & {0.2033} & 0.9448 & 0.9462 & 1.17 & 3.23 \\
        CameraCtrl \citep{he2024cameractrl} & {426.8} & 0.2734 &{0.9526} & {0.9750} & {2.68} & {0.98} 
                    & {675.5} & 0.2027 & {0.9518} & {0.9610} & {1.01} & {4.00} \\
        AC3D~\citep{bahmani2025ac3d} &{415.6} &{0.3044} & {0.9496} &{0.9811} &{2.67} &{0.38} 
             &{452.8} &{0.2078} &{0.9529} & {0.9869} &{1.06} &{2.11} \\
        DualCamCtrl~\citep{zhang2025dualcamctrl} & {408.1} & {0.3154} & {0.9540} & {0.9918} & {1.23} & {0.25} 
                       & {427.4} & {0.2090} & {0.9544} &{0.9807} & {0.83} & {1.53} \\
        \bottomrule
    \end{tabular}}
    \label{table:benchmarkcomp_t2v_combined}
\end{table*}

\subsubsection{Motion-Guided Generation}
Motion-guided video generation is the task of generating videos where a predefined motion concept—such as walking, dancing, waving, or jumping—is explicitly provided as input and used to guide the synthesized motion in the output video. This motion concept represents a coherent and semantically meaningful behavior, typically extracted from one or more reference video clips or motion representations. Unlike standard text-to-video generation that freely imagines motion from prompts, motion-guided methods aim to transfer concrete motion trajectories or dynamics—often in the form of pose sequences, optical flow, or velocity fields—into newly generated visual content. This enables fine-grained control of subject behavior in applications such as character animation, avatar customization, and creative video editing.

The work generally utilizes an explicit motion signal as the control anchor, with an optional text prompt defining scene semantics. Motion control signals vary in format: structured pose sequences like MoTrans~\citep{li2024motrans} and MotionPrompting~\citep{geng2024motionprompting}, hand mask sequences like InterDyn~\citep{akkerman2025interdyn}, or optical flow fields like OnlyFlow~\citep{koroglu2024onlyflow} provide dense or sparse motion descriptors. Some works directly use reference videos as motion priors, including DreamMotion~\citep{jeong2024dreammotion}, DreamRelation~\citep{wei2025dreamrelation}, Customize-A-Video~\citep{ren2024customizeavideo}, DualReal~\citep{wang2025dualrealadaptivejointtraining}, and VideoMage~\citep{Huang_2025_CVPR}. Other works encode abstract fields, such as velocity maps~\citep{liao2025motionagent} or relational cues. Even in motion-guided settings, text prompts are retained to define the appearance, layout, or objects involved in the scene. Furthermore, a growing subset of works supports zero-shot or training-free transfer, including MotionClone~\citep{ling2024motionclonetrainingfreemotioncloning} and VMC~\citep{jeong2023vmc}, by leveraging alignment in latent spaces or score distillation techniques.

To realize controllable generation, a range of architectural designs and mechanisms have been introduced, often combining motion-aware modules with large-scale diffusion backbones. 
DreamActor-M1~\citep{luo2025dreamactor} injects motion tokens into DiT transformers for coherent spatiotemporal control. InterDyn~\citep{akkerman2025interdyn} performs targeted gesture synthesis via attention-guided injection of hand masks. DreamMotion~\citep{jeong2024dreammotion} distills score maps from reference videos to enforce temporal behavior. DreamRelation~\citep{wei2025dreamrelation} models trajectory consistency using relation fields between appearance regions. Methods like MotionAgent~\citep{liao2025motionagent} and MotionPrompting~\citep{geng2024motionprompting} formulate motion as fields or anchor prompts for fine control, while SMA~\citep{pondaven2024spectralmotion} and Diffusion as Shader~\citep{gu2025diffusionasshader} propose spectral and 3D-aware conditioning mechanisms. Meanwhile, MotionMatcher~\citep{wu2025motionmatcher}, Motionbooth~\citep{wu2024motionbooth}, and MotionDirector~\citep{zhao2023motiondirector} align motion tokens in latent space, and Customize-A-Video~\citep{ren2024customizeavideo} learns a one-shot controller from minimal supervision.

Despite these advances, motion-guided generation still faces significant challenges. Motion signals can be noisy, ambiguous, or misaligned with the model's internal representation, often causing motion artifacts or flicker. Recent work MotionPro~\citep{Zhang_2025_CVPR} proposes a region-wise trajectory and motion mask to alleviate misinterpretation and achieve better motion control. 

Furthermore, generalization to unseen motions or subjects remains limited, particularly in one-shot or zero-shot settings (\textit{e.g.}, MotionClone~\citep{ling2024motionclonetrainingfreemotioncloning}, VMC~\citep{jeong2023vmc}). Evaluation also lags behind: although SST-EM~\citep{biyyala2025sstem} introduces metrics for semantic, spatial, and temporal fidelity, widely accepted benchmarks and interpretability tools are under development.

\subsubsection{Summary} Temporal Control activates the dynamic layer of the framework, employing motion modules (\textit{e.g.}, Trajectory Samplers) and camera control modules to translate static concepts into coherent narratives. This modality is central to satisfying the user needs for automation and augmentation, as it replaces manual keyframing with semantic guidance like flow fields or drag-based interactions (\textit{e.g.}, DragAnything~\citep{wu2024draganything}), thereby streamlining the creation of complex scenes.

Regarding the architecture, while UNet models (\textit{e.g.}, Motion-I2V~\citep{10.1145/3641519.3657497}) excel at short-term pixel-level flow dynamics, the complexity of Video-to-4D generation is necessitating a move towards DiT (\textit{e.g.}, DreamActor-M1~\citep{luo2025dreamactor}) and AR backbones. These architectures are favored not just for size, but for their ability to model long-term causal dependencies and 3D geometric consistency (\textit{e.g.}, Voyager~\citep{huang2025voyagerlongrangeworldconsistentvideo}), which are critical when coordinating sophisticated camera paths with object interactions over extended durations.

In the future, the trend of temporal control is moving towards integrated control where temporal signals (Trajectories/Flow) need to coexist with spatial controls (Structure/ID). Current general control models often struggle to merge these distinct signals seamlessly, highlighting a user demand for better interactive editing tools (\textit{e.g.}, drag-based interfaces like DragAnything~\citep{wu2024draganything}) that provide real-time preview of motion paths before generation. Future frameworks will likely prioritize universal control mechanisms to unify these diverse motion inputs into a single interface.

\begin{table*}[t]
\centering
\small
\caption{\textbf{Quantitative comparison with state-of-the-art methods averaging on ambient sound-video, speech-video, and complex scene-video generation.}$\uparrow$ indicates higher is better, $\downarrow$ indicates lower is better. }
\label{tab:joint_comparison}
\resizebox{\linewidth}{!}{
    \renewcommand{\arraystretch}{1.2}
    \begin{tabular}{l|ccccc|cccccc|cccc}
    \toprule
    \multirow{2}{*}{\textbf{Method}} & \multicolumn{5}{c|}{\textbf{Video Quality \& Coherence}} & \multicolumn{6}{c|}{\textbf{Audio Fidelity \& Quality}} & \multicolumn{4}{c}{\textbf{Audio-Visual Synchronization}} \\
    \cmidrule(lr){2-6} \cmidrule(lr){7-12} \cmidrule(lr){13-16}
     & AQ$~\uparrow$ & IQ$~\uparrow$ & DD$~\uparrow$ & MS$~\uparrow$ & ID$~\uparrow$ &  PQ$~\uparrow$ & PC$~\downarrow$ & CE$~\uparrow$ & CU$~\uparrow$ & WER$~\downarrow$ & IB-A$~\uparrow$ & Sync-C$~\uparrow$ & Sync-D$~\downarrow$ & DeSync$~\downarrow$ &  IB$~\uparrow$ \\
    \midrule

     MM-Diffusion~\citep{ruan2023mm} & 0.32 & 0.43 & 0.13 & 0.99  & - & 5.37 & 4.07 & 4.27 & \textbf{5.89} & - & - & - & - & - & 0.12 \\
    
    JavisDiT~\citep{liu2025javisdit} & 0.34 & 0.53 & \textbf{0.38} & 0.99 & 0.38  & 5.46 & 2.24 & 3.19 & 4.54 & 1.00 & \textbf{0.14} & 0.89 & 11.62 & 1.13 & \underline{0.18} \\

    UniVerse-1~\citep{wang2025universe} & 0.52 & \textbf{0.67} & 0.24 & 0.99 & 0.89  & 5.52 & \underline{2.13} & 3.63 & 4.84 & \underline{0.24} & 0.07 & 0.97 & 10.71 & \underline{1.10} & 0.12 \\

    Ovi~\citep{low2025ovi}  & \underline{0.57} & \underline{0.65} & 0.34 & 0.99 & \underline{0.90}  & \underline{6.19} & \underline{2.13} & \underline{4.44} & \underline{5.84} & 0.49 & \underline{0.12} & \underline{4.04} & \underline{9.62} & 1.14 & \underline{0.18} \\
      
    Harmony~\citep{hu2025harmony} & \textbf{0.59} & \underline{0.65} & \underline{0.36} & 0.99 & \textbf{0.91}  & \textbf{6.39} & \textbf{2.05} & \textbf{4.73} & 5.67 & \textbf{0.15} & \underline{0.12} & \textbf{5.61} & \textbf{7.53} & \textbf{0.92} & \textbf{0.19}\\
    
    \bottomrule
    \end{tabular}
}
\end{table*}

\subsection{Audio Control}
\label{sec:audio}
Audio control in video generation refers to the ability to synthesize videos from audio signals such as speech, music, or general sound. This form of control enables models to generate temporally aligned and semantically coherent videos based on audio cues. A subdomain is voice control, which focuses on generating talking face videos or personalized portrait animations conditioned on speech input. Another direction is sound control, which utilizes broader forms of audio (\textit{\textit{e.g.}}, music or ambient sound) to guide the generation with more general content beyond facial animation.
\subsubsection{Voice-Guided Generation}
Voice-guided video generation aims to use both audio and image as conditions to create realistic videos, especially for the talking portrait video generation tasks, which is considered one of the most crucial trends in the applications of the human-computer interaction field. The input embeddings of the sound condition sequences $S_t$ firstly are extracted commonly by the pretrained encoder like wav2vec \citep{schneider2019wav2vecunsupervisedpretrainingspeech}. In order to acquire richer semantic information from the input audio, several works~\citep{xu2024hallohierarchicalaudiodrivenvisual,wang2024vexpressconditionaldropoutprogressive,tian2024emoemoteportraitalive} have concatenated features from different encoder layers for the $f$-th
 frame $S^f$ and injected them into the diffusion processes by the cross-attention layers.

The core issue in talking portrait video generation tasks is capturing and estimating the actual face expressions from the various voice inputs, such as the movements of lips, head poses, and eyebrows, which influence the quality of the video significantly. 
Recently, ~\citep{sun2023vividtalk},~\citep{wei2024aniportraitaudiodrivensynthesisphotorealistic} propose a two-stage architecture based on 3D mesh as the intermediate representation. EDTalk~\citep{tan2024edtalkefficientdisentanglementemotional} has disentangled the motion space into three distinct latent spaces and designs an audio-to-motion module to predict different expressions.  
EMO~\citep{tian2024emoemoteportraitalive} designs the audio feature extractor, face locator, and speed layers, combining with the ReferenceNet~\citep{hu2024animateanyoneconsistentcontrollable} and motion frame module, which ensures the performance of generated videos. EchoMimic~\citep{chen2024echomimiclifelikeaudiodrivenportrait} and EchoMimicV2~\citep{meng2024echomimicv2strikingsimplifiedsemibody} also use the ReferenceNet to extract the features from the input image, generating a better portrait video. 

Different from other conditions like images, layouts, and sketches, the strength of sound is weaker, hindering the control effectiveness during the inference process. Therefore, addressing the conflict between various conditions is another challenge for voice-guided video generation. V-Express~\citep{wang2024vexpressconditionaldropoutprogressive} proposes a progressive training strategy and conditional dropout operations, which are beneficial for those weak control conditions. MOFA-Video~\citep{niu2024mofavideo} designs a MOFA-Adapter to animate a reference image into a video under various control conditions. To balance the control strength between different conditions, MegActor-$\Sigma$ ~\citep{yang2024megactorsigmaunlockingflexiblemixedmodal} proposes a mixed-modal DiT, ACTalker~\citep{hong2025audiovisualcontrolledvideodiffusion} introduced a mask-drop strategy, while MotionCraft~\citep{bian2025motioncraft} and OmniHuman-1~\citep{lin2025omnihuman1rethinkingscalinguponestage} optimize the training strategy of the multimodal generation framework. 

Besides those challenges, there are still some issues, such as limitations in style variation generation on efficiency. For the former one, SVP~\citep{tan2024svpstyleenhancedvividportrait} models the intrinsic style of the source video to alleviate this problem. For the latter one, ~\citep{ki2024floatgenerativemotionlatent, xu2024vasa1lifelikeaudiodriventalking, lin2025cyberhost} improve the real-time performance for the video-chat application. Finally, as for the consistency in long-term video generation, MCDM~\citep{shen2025longtermtalkingfacegenerationmotionprior} designs two clip-motion-prior modules and a memory-efficient attention mechanism. In contrast, TexTalker~\citep{li2025highfidelity3dtalkingavatar} focuses on the dynamic texture for the high-fidelity talking heads generation.
\subsubsection{Sound-Guided Generation}
Sound-guided video generation involves creating videos where visual elements are synchronized with general sound inputs like music, speech, or ambient noises. This method leverages temporal and emotional properties of the sound to guide the visual content, creating videos that reflect not just the overall semantics of the audio but also its dynamic features. In this approach, sound serves as a critical modality for determining the motion, scene structure, and pace of the generated video.

In sound-guided video generation, sound inputs range from specific sounds like animal calls and music to general audio cues. Various audio feature extraction methods, such as Mel frequency cepstral coefficients (MFCCs), Mel spectrograms, BEATs, and CLAP, are used to capture texture, pitch, rhythm, and emotional tones. For instance, AV-Link~\citep{avlink} and ASVA~\citep{Zhang2024AudioSynchronizedVA} synchronize visuals with audio by extracting rhythmic, spectral, and emotional properties using BEATs, CLAP, and MFCCs or Mel-Spectrograms. TA2V~\citep{zhao2024ta2v} further integrates text, using BEATs to guide video generation, showcasing how audio dynamics shape visual output.

Recent advancements in sound-guided video generation, such as AV-Link~\citep{avlink} and ASVA~\citep{Zhang2024AudioSynchronizedVA}, have shifted toward diffusion models for more accurate synchronization, using temporally-aligned activations and audio features to ensure better alignment. TA2V~\citep{zhao2024ta2v} further enhances this by incorporating both text and audio inputs, conditioning video motion and emotion. MotionCraft~\citep{bian2025motioncraft} introduces a dual-branch architecture for rhythmic and semantic alignment, while DAA2V~\citep{yariv2024diverse} refines the process with text-to-video models for improved temporal and semantic alignment. 

Challenges in sound-guided video generation include maintaining synchronization between audio and video, especially with diverse sound types. Solutions like AV-Link\citep{avlink} and ASVA\citep{Zhang2024AudioSynchronizedVA} use temporal attention mechanisms to address this. TA2V\citep{zhao2024ta2v} tackles the complexity of coordinating both text and audio, while DAA2V\citep{yariv2024diverse} provides insights on maintaining synchronization across different audio types.

\subsubsection{Summary} 


In this domain, audio functions as a temporal-dynamic anchor, guiding the rhythmic progression of visual content. A primary challenge is fine-grained temporal alignment to prevent synchronization drift.  This issue has driven a shift from UNet-based pipelines (\textit{\textit{e.g.}}, \textit{EchoMimic}~\citep{chen2024echomimiclifelikeaudiodrivenportrait}) to unified joint generation. Spearheading this evolution, recent works like \textit{MegActor-$\Sigma$}~\citep{yang2024megactorsigmaunlockingflexiblemixedmodal} leverage Diffusion Transformers (DiT) to ensure superior long-range coherence.

Driven by users' demand for high-fidelity interaction, current research tackles a critical control imbalance, where dominant visual features frequently suppress subtle audio cues. It necessitates balanced training strategies, such as the conditional dropout in \textit{V-Express}~\citep{wang2024vexpressconditionaldropoutprogressive}. However, the separation of structural voice and semantic sound control remains a limitation, suggesting a future move toward holistic audio-visual physics to unify micro and macro dynamics.

\subsection{Other Control}
\label{sec:other}
Other control in video generation refers to a set of control mechanisms beyond identity, structure, and motion, enabling more diverse and fine-grained alignment between user intent and generated content. This includes text rendering, which guides generation to produce videos aligned with specific textual elements; style control, which transfers visual characteristics from a reference image or artistic domain to the synthesized video; point control, which enables spatial manipulation through user-specified keypoints or dragging operations; and BEV control, which is especially relevant in autonomous driving scenarios where bird’s-eye view maps or semantic layouts are used to guide video synthesis. 
\subsubsection{Text Rendering}

Text rendering in the video domain targets synthesizing vivid video with aligned text, which has widespread applications in various forms, including advertisements and movies. Previous works~\citep{jiang2023videobooth, shi2024fonts, shi2025wordcon, zeng2023makepixelsdancehighdynamic} leverage the CLIP~\citep{CLIP} text encoder to encode prompts and use 2D VAE~\citep{jiang2025vace} to compress the video from RGB to latent space. They fail to generate clear and aligned text. Text-Animator~\citep{liu2024textanimatorcontrollablevisualtext} designs a text embedding injection module, a camera control module, and a text refinement module to improve both visual text structures and stability. Recently, Wan~\citep{wan2025} has proposed a series of powerful foundation models, including a 1.3B model and a 14B model, which expand the application to various areas such as T2V, text rendering, audio generation, and numerous other downstream tasks.

\subsubsection{Style-Guided Generation}
\label{sec:method_style_guided}
Style-guided video generation transfers an artistic style from a reference image or video to a target video, which is central for creative expression and for achieving specific visual aesthetics. 
The field has progressed from early per-frame stylization with post-hoc temporal smoothing~\citep{chen2017coherent, huang2017real} to efficient and flexible frameworks for arbitrary style transfer~\citep{song2025omniconsistency, guo2025any2anytryon, song2024diffsim, song2024processpainter, zhang2025stable, chen2025transanimate, zhang2024stable, deng2021mccnet, gao2020fast}. 
With the development of generative models, GAN-based approaches support specialized applications such as video toonification~\citep{yang2022vtoonify} and video translation~\citep{yang2023styleganex}, while more recent diffusion models, including Rerender A Video~\citep{yang2023rerender} and StyleCrafter~\citep{liu2023stylecrafter}, now form the leading approach for high-fidelity video stylization.
Building on this, FRESCO~\citep{yang2024fresco} and StyleMaster~\citep{ye2024stylemaster} introduce more fine-grained stylization controls, while UniVST~\citep{song2024univst} aims at a general framework for diverse artistic styles. 
Remaining challenges include temporal consistency, preservation of content details under strong stylization, and intuitive user controls.

\subsubsection{Point-Guided Generation}
\label{sec:point_control}
Point-guided video generation provides fine-grained control by using sparse user-defined point trajectories to animate specific objects or regions, and the main challenge is to extend these constraints in a stable way over space and time. 
GAN-based methods such as Point-to-Point Video Generation~\citep{wang2019point} and ImaGINator~\citep{wang2020imaginator} first explore sparse point inputs for controllable video synthesis, while later work often conditions pre-trained diffusion models on such inputs. 
Track4Gen~\citep{jeong2024track4gen} focuses on robustly tracking user-specified points so that the generated content follows the desired trajectories, and related frameworks such as Diffusion as Shader~\citep{gu2025diffusionasshader} and GS-DiT~\citep{bian2025gs} show that sparse user input can steer more complex scene properties. 
These methods typically encode points as heatmaps or use them to guide attention mechanisms, translating sparse input into coherent and dynamic outputs. Key challenges include physically plausible deformations, long-term temporal consistency from minimal input, and handling complex object interactions and occlusions.

\subsubsection{BEV-Guided Generation}
\label{sec:method_bev_guided}
Bird's-Eye View (BEV) guided video generation targets controllable and realistic driving scenarios for autonomous systems, where a sequence of top-down semantic maps \( B = \{b_1, b_2, \dots, b_T\} \) specifies scene layout and object dynamics. 
Early GAN-based methods such as DriveGAN~\citep{kim2021drivegan} study BEV-conditioned video synthesis for realistic driving sequences, and recent work establishes diffusion models as a central paradigm for translating abstract BEV representations into photorealistic videos, as illustrated by MagicDrive~\citep{gao2023magicdrive}, MagicDrive-V2~\citep{gao2024magicdrivedit}, DriveDreamer-2~\citep{zhao2025drivedreamer}, DiVE~\citep{jiang2024dive}, Delphi~\citep{ma2024unleashing}, DreamForge~\citep{mei2024dreamforge}, Panacea~\citep{wen2024panacea}, and Seeing Beyond Views~\citep{lu2024seeing}. 
Despite this progress, key challenges remain in geometric consistency between BEV inputs and rendered videos, realistic appearance and lighting, accurate modeling of complex dynamic interactions, and the semantic gap between abstract BEV representations and fine-grained visual details.

\subsubsection{Summary} Other control modalities, including text rendering, style transfer, point tracking, and BEV conditioning, provide fine-grained and semantically aligned control over video generation beyond identity, structure, and motion. 
In our framework, these modalities connect foundation models with high-level creative tools through conditional encoders and control modules such as adapters and cross-attention mechanisms, which inject sparse or abstract guidance into the generative process. 
The design trend is shifting from mainly UNet-based architectures (for example, ControlNet-based stylization) toward DiT-based backbones, as in Wan and advanced BEV models, to better capture global semantic context and complex spatio-temporal dependencies for tasks such as coherent text spelling and autonomous driving simulation. 
From a user perspective, these controls support personalization through style and text and real-time preview and adjustment through point dragging, while turning these mechanisms into intuitive, transparent tools with effective automatic suggestions remains an important direction for future research.

\subsection{Universal Control}
\label{sec:universal}
Universal Control refers to the ability to flexibly manage and integrate various types of input conditions—such as text, spatial features, and temporal information—to generate videos that meet specific user requirements. This method allows for high customization, enabling users to influence multiple aspects of the generated video simultaneously, such as object behavior, motion, camera angles, and scene layout. By unifying different conditions into a single, coherent framework, universal control facilitates comprehensive video synthesis with enhanced flexibility and precision.

In universal-guided video generation, the model typically incorporates a range of inputs, including text, spatial conditions (such as camera poses, depth maps, and human poses), and temporal signals (including motion sequences and video clips). For example, frameworks like VideoComposer \citep{wang2023videocomposer} utilize motion vectors or sketches as explicit control signals, guiding the video generation process. Similarly, Any2Caption \citep{wu2025any2caption} employs multimodal inputs, such as text, camera poses, and human motions, which are interpreted into structured captions to drive video synthesis. FullDiT \citep{ju2025fulldit} integrates these various conditions using a unified attention mechanism, processing text, camera views, depth information, and identity conditions simultaneously, thus enabling seamless video generation.


Several recent frameworks highlight different ways to achieve this. For instance, VideoComposer~\citep{wang2023videocomposer} relies on a dedicated spatio-temporal condition encoder to blend motion cues, depth information, and text prompts inside a latent diffusion pipeline. Any2Caption~\citep{wu2025any2caption}, on the other hand, takes a more indirect route: it first translates diverse inputs into detailed, structured textual captions using a multimodal LLM, then feeds those captions to a standard video generator. FullDiT~\citep{ju2025fulldit} pushes direct integration further by employing full spatiotemporal attention within a Diffusion Transformer architecture, letting it handle text, camera poses, depth, and even identity constraints all at once. Notably, in the autoregressive camp, InfinityStar~\citep{liu2025infinitystar} stands out—it simply tokenizes everything (visual content and controls alike) into a unified discrete sequence, enabling native multi-condition control for tasks ranging from text-to-video to image animation and long-sequence extension, all without extra adapters.

Despite the progress made in universal-guided control, several challenges persist. One of the primary difficulties lies in balancing the integration of various conditions. Ensuring that each condition (\textit{e.g.}, motion, text, camera) is harmoniously combined without one overshadowing the others remains a complex task. In addition, maintaining temporal consistency across frames presents another challenge, particularly when dealing with dynamic and varied motion patterns. Furthermore, evaluation continues to be an unresolved issue—while benchmarks like FullBench for FullDiT have been introduced, the development of comprehensive and standardized evaluation criteria for universal control across all modalities is still underway.

In summary, universal control uses advanced conditioning—from encoders and attention fusion to token unification—to combine signals like text, poses, depth, and motion. This lets users customize content, style, motion, and layout all at once, far surpassing what single-condition models can do.

The architecture has shifted from bulky UNet-with-adapters designs to more scalable backbones, driven by the need for cleaner multi-condition handling and better scaling. Diffusion Transformers (\textit{e.g.}, FullDiT~\citep{ju2025fulldit}) and autoregressive models (\textit{e.g.}, InfinityStar~\citep{liu2025infinitystar}) now lead, offering natural fusion of controls, support for long/high-res sequences, and either high fidelity or fast in-context flexibility. Caption approaches like Any2Caption~\citep{wu2025any2caption} and early composable encoders from VideoComposer~\citep{wang2023videocomposer} remain useful, but trends favor integrated, interactive systems. Challenges in signal conflicts, long-term coherence, and interpretability persist, suggesting future progress will focus on hybrid designs and clearer user insight into control interactions.

\begin{figure*}[t]
    \centering
    \includegraphics[width=1.0\linewidth]{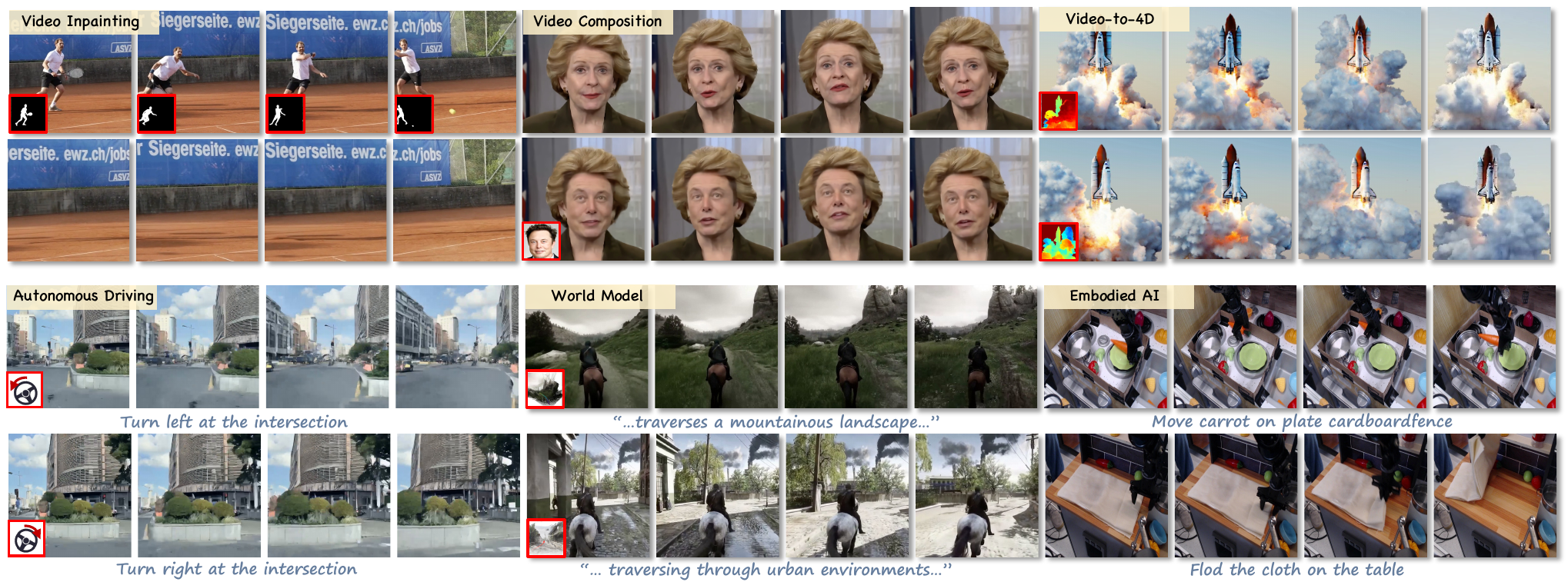}
    \caption{\textbf{Applications of Controllable Video Generation.} The sample images are from DiffuEraser~\citep{li2025diffueraserdiffusionmodelvideo}, VideoAnydoor~\citep{tu2025videoanydoorhighfidelityvideoobject}, 4D-fy~\citep{4d-fy}, Vista~\citep{gao2024Vista}, spmem~\citep{wu2025spmem}, RoboMaster~\citep{fu2025learningvideogenerationrobotic}.}
    \label{fig:Application}
\end{figure*}

\section{Application}
\label{sec:Application}

With the rapid advancement of video generative models, video generation techniques are increasingly being applied to real-world scenarios. As illustrated in Fig.~\ref{fig:Application}, T2V models have demonstrated significant potential in various areas.

\noindent \textbf{Video Completion and Inpainting.} Video completion and inpainting aim to fill in missing areas in videos or remove unnecessary objects while maintaining spatial and temporal consistency. The techniques can be widely used in film post-production, historical video restoration, and everyday mobile video editing apps. Traditional methods often struggle with large occlusions or complex motion. In recent years, video generative models have shown strong generative capabilities in this field. DiffuEraser~\citep{li2025diffueraserdiffusionmodelvideo} is designed to fill occluded areas, combining prior information for initialization and weak conditional constraints to help reduce noise artifacts and suppress hallucination. AVID~\citep{zhang2024avidanylengthvideoinpainting} introduces motion modules and adjustable structural guidance, which support different types of repair under various levels of structural fidelity, capable of handling videos of any length, ensuring temporal consistency within the editing area.

\noindent \textbf{Video Composition.} Video composition aims to seamlessly embed target objects into existing videos while ensuring both temporal and spatial consistency. It is essential in advertising production, AR/VR environments, and creator tools such as short-video platforms. It allows users to insert virtual objects into real scenes, enabling applications like virtual product placement, AR furniture try-on, or creative storytelling with animated characters. It significantly reduces the effort needed to produce high-quality mixed-reality content. This task requires the inserted object to be well integrated with the background while preserving the details of the appearance of object. Recent advances in text-to-video diffusion models offer powerful control capabilities for this purpose. For example, MVOC~\citep{wang2024mvoctrainingfreemultiplevideo} reverses the corresponding noise features by DDIM inversion on each video object, then synthesizes and edits them to generate the first frame of the synthetic video. VideoAnydoor~\citep{tu2025videoanydoorhighfidelityvideoobject} is an end-to-end video object insertion framework that achieves high-fidelity detail retention and precise motion control through the ID extractor, Pixel Warper, and optimized training strategies.

\noindent \textbf{Video-to-4D Generation.} 4D generation is inherently more challenging than static 3D reconstruction, as it requires modeling both spatial geometry and temporal dynamics. Video generative models naturally provide strong motion priors by capturing scene dynamics over time, making them an emerging core technology for constructing high-quality 4D representations. It enables game studios, film creators, and cultural heritage organizations to build 4D digital assets from simple handheld videos, avoiding expensive multi-camera capture systems. It can also support robotic perception by providing realistic, dynamic 3D environments for simulation and training. 4D-fy~\citep{4d-fy} proposes a mixed score distillation method that combines multiple pre-trained diffusion models to generate 4D scenes with realistic appearance, structure, and motion. 4Real~\citep{4real} utilizes a video diffusion model to generate reference videos and ``freeze-time" videos, and learns a standard 3D representation based on deformable 3D Gaussian splatting and time-varying deformations.

\noindent \textbf{Autonomous Vehicle and Generative World Model.} World model is important in autonomous driving by simulating environments and guiding decision-making. Video generation models, with their ability to produce realistic visual data, are now increasingly used to improve perception and planning in autonomous systems. GAIA-2~\citep{GAIA-2} is a multi-view generative world model capable of producing high-resolution, temporally and spatially consistent multi-camera videos from structured inputs such as vehicle dynamics, environmental conditions, and road semantics. DriVerse~\citep{DriVerse} generates high-fidelity driving simulation videos by using a single image and future trajectories, introduces multi-modal trajectory prompts to encode trajectories as text and spatial motion priors, and uses latent motion alignment to enhance the temporal consistency of dynamic objects. Drive-WM~\citep{wang2023Drivingintothefuture} promotes spatial-temporal joint modeling through view factorization, enabling the generation of high-fidelity multi-view videos in autonomous driving scenarios. The powerful generative capability demonstrates the potential of applying world models to safe driving planning.

\noindent \textbf{Embodied Artificial Intelligence.} Video generative models can simulate future scenarios, enabling embodied agents to understand, predict, and plan interactive behaviors by generating visual information that aligns with semantic and physical constraints. This enhances the agents' capabilities within the perception–decision–action loop. Gen2Act~\citep{Gen2Act} introduces a framework that uses video generation model to generate human operation videos. These videos are translated into executable robot actions via a policy model, addressing the challenge of generalizing to unseen objects and novel tasks. AVDC~\citep{AVDC} leverages video diffusion model to synthesize hallucinated videos of robots performing actions. It extracts dense correspondences via optical flow within the synthesized video and combines them with initial depth estimates, thereby inferring robot actions. RoboMaster~\citep{fu2025learningvideogenerationrobotic} generates robot manipulation videos through collaborative trajectory control. It addresses the feature entanglement problem in multi-object interaction by decomposing the process into three stages and integrating object embeddings with appearance and shape perception. This\&That~\citep{wang2024language} enhances robot planning and execution by generating task videos conditioned on both language and gestures. Compared to language-only methods, this multimodal approach enables more accurate interpretation of user intentions.

\section{Discussion and Future Work}
\label{sec:discussion and future}
Controllable video generation has advanced rapidly (Tab.~\ref{tab:methods}), yet building robust and scalable systems that support diverse controls remains challenging. Current methods often optimize a subset of constraints (\textit{e.g.}, structure, temporal consistency, style), leading to unavoidable trade-offs when multiple conditions must be satisfied simultaneously. We highlight three promising directions for future research.

\subsection{Unified Multi-Condition Control and Constraint Orchestration}
A key bottleneck is compositional controllability: real applications usually require jointly satisfying temporal coherence, spatial/structural fidelity, motion dynamics, and stylistic consistency. Existing works typically emphasize one aspect at a time—\textit{e.g.}, structure-driven generation such as Follow-Your-Pose~\citep{ma2024followposeposeguidedtexttovideo} and DriveDreamer~\citep{zhao2025drivedreamer}, temporal control such as MotionBooth~\citep{wu2024motionbooth} and Direct-A-Video~\citep{yang2024directavideo}, and image/style-guided generation such as VideoCrafter1~\citep{chen2023videocrafter1opendiffusionmodels} and Lumiere~\citep{10.1145/3680528.3687614}. As a result, enforcing one constraint often degrades others.

Future systems should move toward constraint orchestration: (i) hierarchical or modular control frameworks that compose heterogeneous conditions and explicitly manage conflicts; (ii) adaptive control policies that can re-weight constraints based on user intent and content (\textit{e.g.}, prioritizing temporal smoothness for action sequences while emphasizing style for artistic videos). VideoComposer~\citep{wang2023videocomposer} is an early step toward universal conditioning, but stronger mechanisms are needed to reliably handle complex, mixed user inputs. Compositional diffusion and unified architectures (\textit{e.g.}, FullDiT~\citep{ju2025fulldit}) further suggest the feasibility of multi-condition integration, yet open problems remain in balancing competing constraints and maintaining computational efficiency.

\subsection{Unified Video Reasoning and Generation}
Another direction is to couple reasoning with generation~\citep{chen2025videochat,chen2025top,chen2025g}. LLMs can serve as a semantic control layer that translates high-level intent into concrete, low-level conditions and supports iterative refinement through multi-turn interaction. Recent systems such as Phantom~\citep{liu2025phantom} demonstrate progress toward semantically rich, prompt-driven generation, but robust alignment between reasoning modules and video generators is still limited, especially under complex instructions and long-horizon consistency requirements.

A promising route is multi-modal alignment that allows users to combine text with other signals while keeping the interaction natural and editable. In this context, Multi-Modal LLM (MLLM) agents can act as a unified interface for planning, grounding, and refinement: they can parse intent, select or synthesize appropriate controls, and iteratively update constraints based on user feedback. Key challenges include (i) reliable grounding and verification to prevent instruction drift and hallucinated content; (ii) bias and safety mitigation in open-ended generation; and (iii) efficiency, \textit{e.g.}, lightweight adapters or cross-modal modules that reduce compute overhead while preserving controllability.

\subsection{Hybrid and Scalable Autoregressive Methods}
Scaling to long videos with strong temporal coherence remains difficult. Diffusion models produce high-quality frames but may suffer from accumulated inconsistencies over long durations, whereas autoregressive generation explicitly models temporal dependencies and can improve coherence (\textit{e.g.}, NOVA~\citep{deng2025autoregressivevideogenerationvector}), at the cost of higher latency and compute.

Hybrid designs that combine diffusion and autoregressive components are therefore attractive: diffusion can generate globally consistent latent representations, while lightweight autoregressive modules enforce sequential continuity. Multi-scale modeling is also promising, capturing coarse dynamics at low temporal/spatial resolution and refining details progressively (\textit{e.g.}, Motion-I2V~\citep{10.1145/3641519.3657497}). For further scalability, memory-efficient architectures (\textit{e.g.}, sparse attention, recurrent latent modules) can reduce long-range modeling cost, and large-scale self-supervised pretraining on datasets such as WebVid-10M~\citep{Bain21WebVid-10M} and Panda-70M~\citep{chen2024panda-70m} can improve generalization and reduce reliance on task-specific tuning.

Overall, advances in hybrid architectures and efficient long-range modeling are likely to be central to practical, high-quality, and controllable long-video generation for applications such as storytelling, simulation, and animation.
\section{Conclusion}

This paper presents a comprehensive survey of controllable video generation based on foundational generative models. First, we introduce the core theoretical foundations, including GANs, VAEs, denoising diffusion probabilistic models, flow-based models, and autoregressive architectures, along with representative video generative models. Then, we propose a structured taxonomy that categorizes existing controllable generation methods by their conditioning signals beyond text.
Next, we review representative techniques for incorporating novel conditions into video generation pipelines, tracing their development across different theoretical paradigms and model designs. 
We further synthesize prior research by examining its progression from core principles to technical innovations and implementation strategies. Additionally, we highlight real-world applications where controllable video generation has demonstrated its practical impact, emphasizing its relevance and future promise within the broader AIGC ecosystem. Through this survey, we aim to present a holistic understanding of the field’s current landscape and provide insights that inform future directions in controllable video synthesis research. 

\noindent\textbf{Data Availability Statements.} All the data used in this paper can be downloaded from the Internet, and the authorization has been obtained, including civitai~\citep{civitai}. 

\bibliography{main}

\end{document}